\documentclass{aastex631}

\usepackage{bm}

\shortauthors{Wang \& Yang}

\graphicspath{{./}{figures/}}

\begin{document}

\title{Lorenz Energy Cycle: Another Way to Understand the Atmospheric Circulation on Tidally Locked Terrestrial Planets}

\author[0000-0002-9406-1781]{Shuang Wang}
\affiliation{Laboratory for Climate and Ocean-Atmosphere Studies, Dept. of Atmospheric and Oceanic Sciences, School of Physics, Peking University, Beijing 100871, China}

\author[0000-0001-6031-2485]{Jun Yang}
\affiliation{Laboratory for Climate and Ocean-Atmosphere Studies, Dept. of Atmospheric and Oceanic Sciences, School of Physics, Peking University, Beijing 100871, China}

\correspondingauthor{Jun Yang}
\email{junyang@pku.edu.cn}


\begin{abstract}
In this study, we employ and modify the Lorenz energy cycle (LEC) framework as another way to understand the atmospheric circulation on tidally locked terrestrial planets. It well describes the atmospheric general circulation in the perspective of energy transformation, involved with several dynamical processes. We find that on rapidly rotating, tidally locked terrestrial planets, mean potential energy (P$_{\rm M}$) and eddy potential energy (P$_{\rm E}$) are comparable to those on Earth, as they have similar steep meridional temperature gradients. Mean kinetic energy (K$_{\rm M}$) and eddy kinetic energy (K$_{\rm E}$) are larger than those on Earth, related to stronger winds. The two conversion paths, P$_{\rm M}\rightarrow$P$_{\rm E}\rightarrow$K$_{\rm E}$ and P$_{\rm M}\rightarrow$K$_{\rm M}\rightarrow$K$_{\rm E}$, are both efficient. The former is associated with strong baroclinic instabilities, and the latter is associated with Hadley cells. On slowly rotating, tidally locked terrestrial planets, weak temperature gradients in the free atmosphere and strong nightside temperature inversion make P$_{\rm M}$ and P$_{\rm E}$ are much smaller than those on Earth. Meanwhile, large day--night surface temperature contrast and small rotation rate make the overturning circulation extend to the globe, so that the main conversion path is P$_{\rm M}\rightarrow$K$_{\rm M}\rightarrow$K$_{\rm E}$. This study shows that the LEC analyses improve the understanding of the atmospheric circulation on tidally locked terrestrial planets.    
 
\end{abstract}


\section{Introduction}\label{sec:intro}

The substellar point of 1:1 tidally locked (or synchronously rotating) terrestrial planet is fixed with time. Such a state can drive different atmospheric circulation compared to Earth, which has been simulated by general circulation models (GCMs) in previous studies \citep[e.g.,][]{Joshi_1997,Merlis_2010,Edson_2011,Leconte_2013,Wordsworth_2015,Koll_2016,Noda_2017,Haqq_Misra_2018,Pierrehumbert_Hammond_2019,Hammond_2021,Sergeev_2022,Turbet_2022,Wang_2022}. These simulations showed that the atmospheric circulation on tidally locked terrestrial planets is dominated by a global-scale overturning circulation, consisting of winds with upwelling in the substellar region, \textcolor{black}{horizontally} flowing from the dayside to the nightside in the upper troposphere, downwelling in the region away from the substellar point, and flowing back from the nightside to the dayside near the surface. Besides, there are a westerly jet over the equator (equatorial superrotation) and planet-sized wavenumber-1 stationary Rossby and Kelvin waves.

For understanding the atmospheric circulation on tidally locked terrestrial planets, various methods have been used. Momentum budgets were widely used to explore the interactions between zonal jets and planetary waves \citep[e.g.,][]{Showman_2010,Showman_2011,Perez_Becker_2013,Tsai_2014,Hammond_2018,Mendonca_2020,Debras_2020,Hammond_2020,Wang_2020}. It well demonstrated that up-gradient momentum transports by the stationary waves maintain the equatorial superrotation against friction. This method was also beneficial for the prediction of the equatorial jet speed \citep{Hammond_2020}. Besides, a Helmholtz decomposition technique, which was suggested by \cite{Hammond_2021}, was also used in several recent studies \citep{Ding_2021,Sergeev_2022,Turbet_2022}. This technique separated the total circulation into divergent (overturning circulation) and rotational components (zonal jets and stationary waves). It was helpful to classify the dynamical regimes and to quantify the transports of energy and tracers. For example, \cite{Hammond_2021} found that the global overturning circulation could dominate the day--night heat transport even when the zonal jet is strong.

Another method is describing the atmospheric circulation from the perspective of energy and energy transformation. Atmospheric heat engine framework depicts the Earth's atmospheric circulation as an ideal Carnot's heat engine \citep{Peixoto_1992}. \cite{Koll_2016} applied this framework on the tidally locked terrestrial planet: the atmosphere absorbs heat on the dayside at a dayside surface temperature of $T_{d}$ and emits it to space at a planet's equilibrium temperature of $T_{eq}$, which allows the atmosphere to maintain the global overturning circulation against friction in the boundary layer, following a Carnot's efficiency of $\eta=(T_{d}-T_{eq})/T_{d}$. By using this framework, they estimated the surface wind speed and developed an upper limit on the strength of the overturning circulation.

\begin{figure*}
    \centering
    \includegraphics[width=0.9\textwidth]{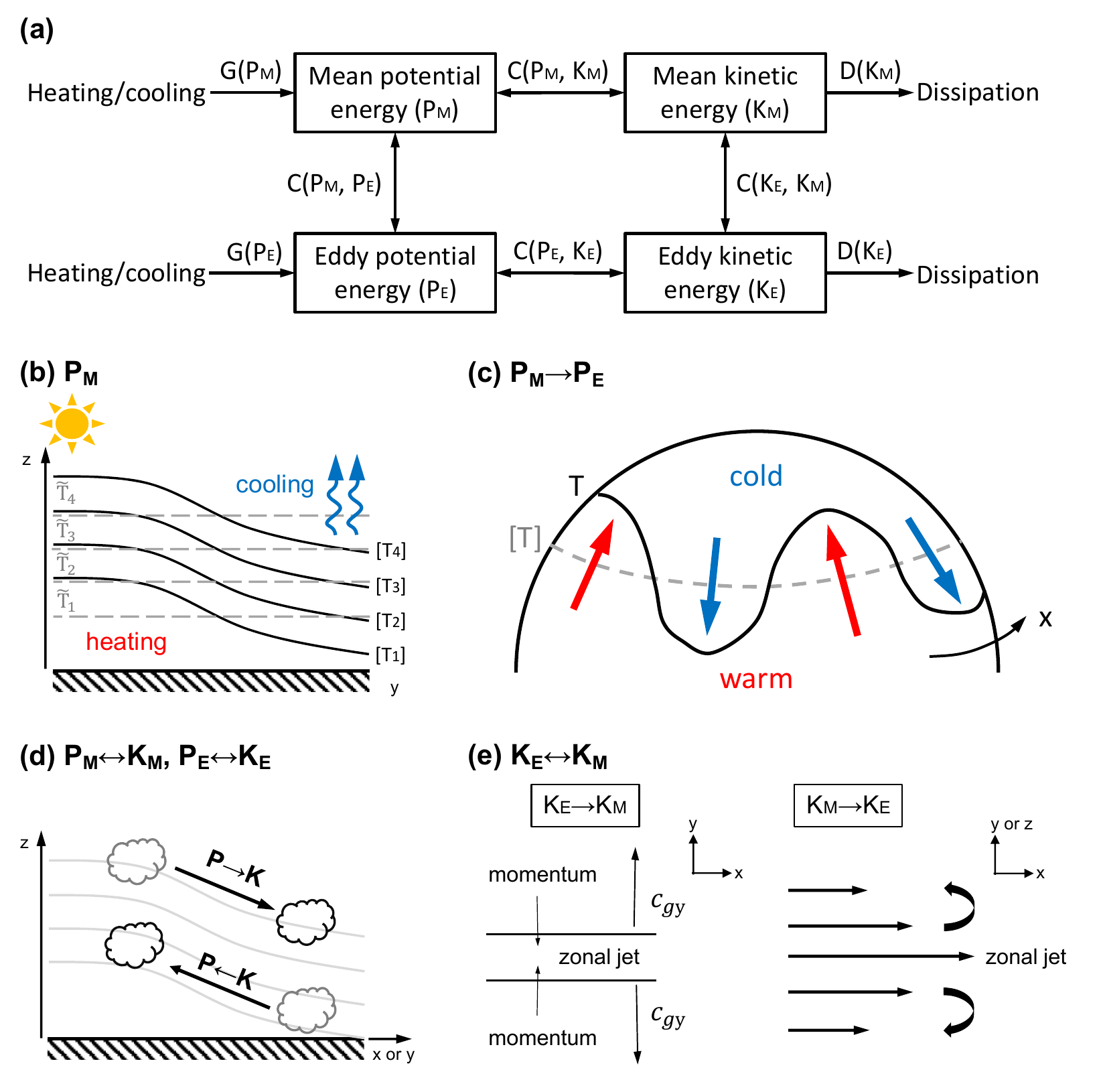}
    \caption{(a) Framework of the Lorenz energy cycle (LEC) on Earth: boxes represent the energy reservoirs; arrows represent generation (G), transformation (C), and dissipation (D) of energy. (b) Generation of P$_{\rm M}$: uneven heating and cooling tilt the zonal-mean isotherms (black curves) departed from the global-mean isotherms (grey dashed lines), which generates P$_{\rm M}$. (c) Heat transport by baroclinic eddies, converting P$_{\rm M}$ to P$_{\rm E}$: the zonal-mean isotherm (grey dashed line) is warped to the isotherm (black curve) by baroclinic eddies, which decreases zonal-mean meridional difference, i.e., P$_{\rm M}$, but leads zonal variance, i.e., P$_{\rm E}$. (d) Cross-isobaric motions, converting potential energy and kinetic energy with each other.  (e) Wave--mean flow interactions, converting K$_{\rm E}$ and K$_{\rm M}$ with each other: \textcolor{black}{under a positive $\beta$-plane, arbitrary waves with group velocity directed away from the source region in the y-direction ($c_{gy}$) can transport momentum back}, along with conversion from K$_{\rm E}$ to K$_{\rm M}$; shear instabilities of zonal jets generate eddies, along with conversion from K$_{\rm M}$ to K$_{\rm E}$.}
    \label{fig:LEC_framework}
\end{figure*}

In this study, the Lorenz energy cycle \citep[LEC,][]{Lorenz_1955} is employed as another way to understand the atmospheric circulation on tidally locked terrestrial planets. It describes the general circulation from the perspective of energy transformation, and has been widely \textcolor{black}{applied to Earth. For example, the LEC has been used to understand the atmospheric circulation and dynamical processes, such as waves, zonal jets, and their interactions \citep[e.g.,][]{Peixoto_1974,Peixoto_1992,Ulbrich_1991,Duan_2005}. Moreover, the ability to simulate LEC calculated from observational data is a useful diagnostic for climate models \citep[e.g.,][]{Deckers_2010,Marques_2011}. Comparison of the LEC calculated from various reanalysis data is also beneficial to evaluate these data \citep[e.g.,][]{Ulbrich_1991,Li_2007,Marques_2009,Marques_2010,Kim_2013}. Recently, the LEC has been applied to show the variability of the energy cycle in response to climate change over the last 40 years \citep[e.g.,][]{Kim_2017,Pan_2017}. The same framework also has been used to predict the future variability of the energy cycle in the Coupled Model Intercomparison Project \citep[e.g.,][]{Michaelides_2021,Yuki_2022}.} Briefly, the net incoming solar radiation, latent heat release in the tropics, and net infrared cooling in mid- and high-latitudes together generate mean potential energy (P$_{\rm M}$) in Earth's atmosphere. The growing baroclinic eddies convert P$_{\rm M}$ to eddy available potential energy (P$_{\rm E}$), and then convert P$_{\rm E}$ to eddy kinetic energy (K$_{\rm E}$). A portion of K$_{\rm E}$ is converted to the mean kinetic energy (K$_{\rm M}$) through wave--mean flow interactions. The bulk of K$_{\rm M}$ and K$_{\rm E}$ is ultimately dissipated through small-scale turbulence and surface friction (Figure~\ref{fig:LEC_framework}). The main pathway of enengy conversion on Earth follows P$_{\rm{M}}\rightarrow$P$_{\rm{E}}\rightarrow$K$_{\rm{E}}\rightarrow$K$_{\rm{M}}$. In addition, the LEC has also been used to understand the oceanic circulation on Earth \citep[e.g.,][]{Peixoto_1992,Olbers_2012,Storch_2012} and the atmospheric circulation on other planets such as Venus and Titan \citep[e.g.,][]{DelGenio_1993,Yamamoto_2006,Lee_2010}. 

A key point of this study is that we apply the LEC to tidally locked terrestrial planets. We evaluate the LEC on Earth-like tidally locked planets and compare it with that on Earth. We find that P$_{\rm M}$ is very small on slowly rotating tidally locked planets, and the main path of the energy conversion is P$_{\rm{M}}\rightarrow$K$_{\rm{M}}\rightarrow$K$_{\rm{E}}$. We also find that the LEC on rapidly rotating planets is like a combination of those on Earth and on slowly rotating planets. The structure of this paper is as follows. Section~\ref{sec:data} describes the methodology and data. Section~\ref{sec:result} shows the thermal structure, atmospheric circulation, and LECs on three different types of planets. Section~\ref{sec:summary} is the summary and discussions.

\section{Methodology and data}\label{sec:data}
\subsection{LEC in tidally locked coordinates}
\begin{figure*}
    \centering
    \includegraphics[width=0.9\textwidth]{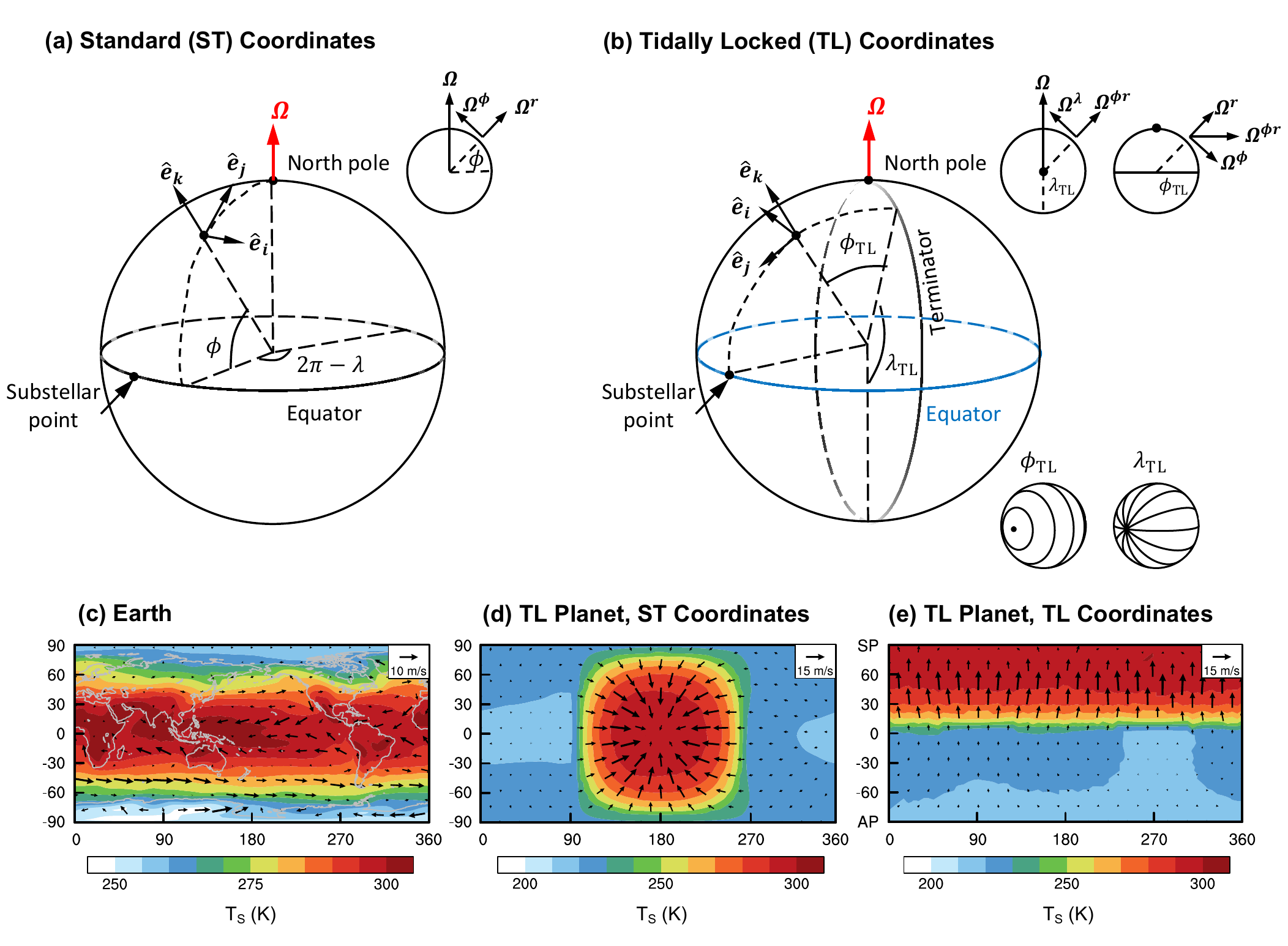}
    \caption{(a) Frame of standard (ST) coordinates. The substellar point is located at latitude/longitude $\left(\phi,\lambda\right)=\left(0^{\circ},180^{\circ}\right)$. Three black arrows are unit vectors of the local Cartesian coordinates $(\mathbf{\hat{e}_i},\mathbf{\hat{e}_j},\mathbf{\hat{e}_k})$. Red vector is the rotational angular velocity, and its projections onto the local axes are shown at the top-right corner. (b) Frame of tidally locked (TL) coordinates. The substellar point is located at tidally locked latitude $\phi_{TL}=90^{\circ}$. The projections of the rotational angular velocity are shown at the top-right corner, and the solid line represents the terminator, and dots represent the substellar points. The frames of the tidally locked latitude and longitude lines are shown at the bottom-right corner. (c) Annual-mean surface temperature (contours) and the near-surface winds (vectors) on Earth. (d) Same as (c) but on a tidally locked terrstrial planet with an orbit period of 60 Earth days. (e) Same as (d) but shown in the tidally locked coordinates. SP and AP are the substellar point and the antistellar point, respectively.}
    \label{fig:changing_coordinates}
\end{figure*}
LEC is more regarded as \textcolor{black}{a fundamental property} of the Eulerian mean system rather than of the real atmosphere, because the Eulerian means are employed to define the mean energy, and the departures from the Eulerian means are employed to define the eddy energy \citep[\textcolor{black}{Chapter 10.4 in}][]{Holton_2013}. Thus, it is important to employ a suitable Eulerian mean. For Earth, zonal means are always employed, because solar radiation, temperature, and winds are almost zonally homogeneous for long-term averages (Figure~\ref{fig:changing_coordinates}(c)). For tidally locked terrestrial planets, especially slowly rotating ones, the Eulerian mean should match the monotonically decreasing stellar radiation and temperature along an arbitrary direction from dayside to nightside (Figure~\ref{fig:changing_coordinates}(d)). So for slowly rotating tidally locked terrestrial planets, the tidally locked coordinates are employed in this study, similar to that used in previous studies \citep[e.g.,][]{Koll_2015,Koll_2016,Ding_2020a,Ding_2020b,Hammond_2021,Sergeev_2022,Turbet_2022,Wang_2022}. In the tidally locked coordinates, the nominal ``North/South Pole'' is the substellar/antistellar point, the tidally locked latitude lines are a series of concentric circles around the substellar point, and the tidally locked longitude lines are the great circles linking the substellar and antistellar points (Figure~\ref{fig:changing_coordinates}(b)). 
The transformation relations between the standard and the tidally locked coordinates are given in Appendix~\ref{appendix_a}. By the transformation, the overturning circulation (Figure~\ref{fig:changing_coordinates}(d)) become a zonal-mean component, i.e., homogeneous along the tidally locked longitudes (Figure~\ref{fig:changing_coordinates}(e)). That is, the Eulerian mean on a tidally locked planet could be defined as the zonal averages in the tidally locked coordinates.

\begin{figure*}
    \centering
    \includegraphics[width=0.9\textwidth]{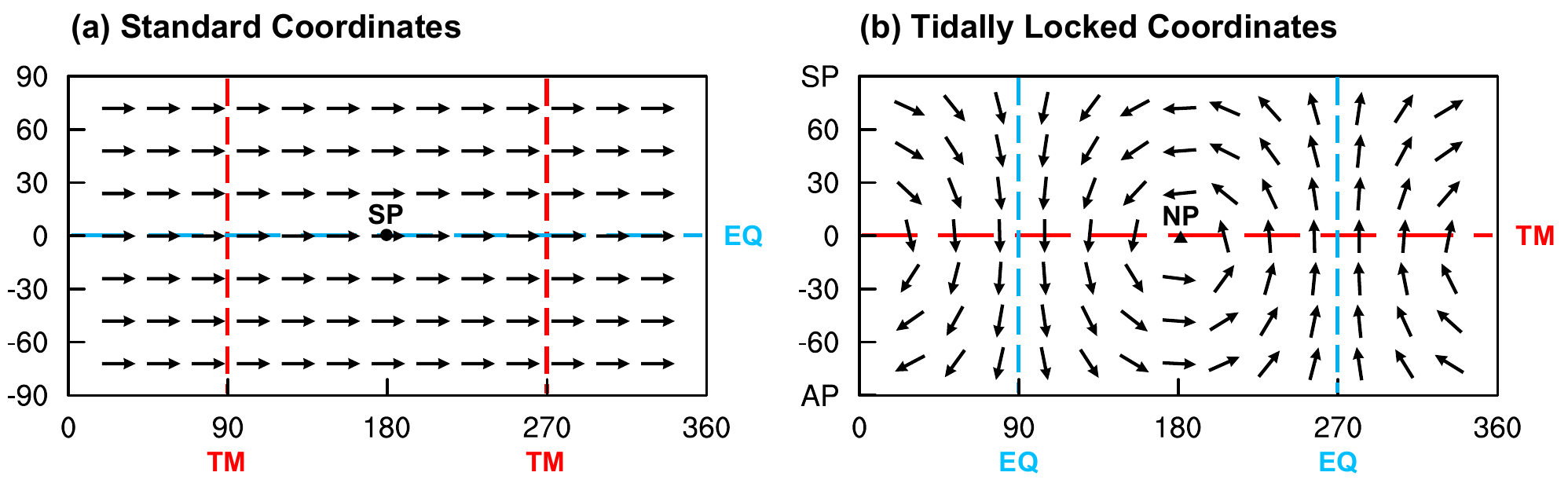}
    \caption{A hypothetical uniform zonal-mean zonal jet in the standard coordinates (a) is transformed into the winds in the tidally locked coordinates (b) using Equations (\ref{eq:special_U0}) and (\ref{eq:special_V0}). Blue dashed lines represent the equator (EQ) in the standard coordinates; red dashed lines represent the terminators (TM); black dot in panel (a) represents the substellar point (SP); black triangle in panel (b) represents the North Pole (NP) in the standard coordinates.}
    \label{fig:transform_zonal_jet}
\end{figure*}

Note that a zonal-mean zonal jet in the standard coordinates would be transformed into an eddy component in the tidally locked coordinates. For example, a uniform zonal-mean zonal jet in the standard coordinates $U_0$, as shown in Figure~\ref{fig:transform_zonal_jet}(a), would be transformed into corresponding winds in the tidally locked coordinates following the relations of  
\begin{eqnarray}
    u_{TL}&=&\frac{\cos\lambda_{TL}\tan\phi_{TL}}{\sqrt{\sin^2\lambda_{TL}+\tan^2\phi_{TL}}}U_0, \label{eq:special_U0}\\
    v_{TL}&=&-\frac{\sin\lambda_{TL}\sec\phi_{TL}}{\sqrt{\sin^2\lambda_{TL}+\tan^2\phi_{TL}}}U_0, \label{eq:special_V0}
\end{eqnarray}
where $u_{TL}$ is tidally locked zonal wind, $v_{TL}$ is tidally locked meridional wind, $\lambda_{TL}$ is tidally locked longitude, and $\phi_{TL}$ is tidally locked latitude \textcolor{black}{(see Equation~(\ref{eq:wind_relations_matrix_tl}) in Appendix~\ref{appendix_a}).} Figure~\ref{fig:transform_zonal_jet}(b) shows the transformed winds from Equations~(\ref{eq:special_U0}) and (\ref{eq:special_V0}), and suggests that the zonal-mean zonal jet becomes an eddy component.

A critical step is deriving the governing equations in the tidally locked coordinates. We start this step from the primitive equations, including a momentum vector equation,
\begin{equation}\label{eq:general_momentum_equation}
    \frac{D\mathbf{v}}{Dt}+2\bm{\Omega}\times\mathbf{v}=-\frac{1}{\rho}\nabla p+\mathbf{g}+\mathbf{F},
\end{equation}
a mass continuity equation,
\begin{equation}\label{eq:continuity}
    \frac{D\rho}{Dt}+\rho\nabla\cdot\mathbf{v}=0,
\end{equation}
and a thermodynamic equation,
\begin{equation}\label{eq:thermodynamic}
    \frac{DT}{Dt}+\frac{p}{\rho c_v}\nabla\cdot\mathbf{v}=\frac{\dot{Q}}{c_v},
\end{equation}
where $D/Dt=\partial/\partial t+\mathbf{v}\cdot\nabla$, $\mathbf{v}\cdot\nabla$ is advection operator, $\mathbf{v}$ is velocity of flows, $\bm{\Omega}$ is planetary rotation rate, $\rho$ is air density, $p$ is pressure, $\mathbf{g}$ is effective gravity vector, $\mathbf{F}$ is friction force, $T$ is air temperature, $c_v$ is specific heat at constant volume, and $\dot{Q}$ is heating rate \citep[\textcolor{black}{Chapter 1 in}][]{Vallis_2019}. Mathematically, the transformation of coordinates does not change the values and forms of scalars and scalar operators, i.e., temperature, density, divergence of velocity, heating rate, and advection operator \citep[Chapter 2 in][]{Kundu_fluid}. Thus, the two scalar equations, Equations~(\ref{eq:continuity}) and (\ref{eq:thermodynamic}), will not change. However, the transformation of coordinates will change the projections of vectors onto new axes, i.e., wind velocity, rotation rate, and pressure gradient, so that the projections of Equation~(\ref{eq:general_momentum_equation}) will change. 

Considering that the directions of unit vectors of axes change as these vectors move with the atmosphere, it would introduce an effective rotation rate, so that Equation~(\ref{eq:general_momentum_equation}) in the tidally locked coordinates is written as    
\begin{equation}\label{eq:scalarform_momentum_equation}
    \frac{Du_{TL}}{Dt}\mathbf{\hat{e}_i}+\frac{Dv_{TL}}{Dt}\mathbf{\hat{e}_j}+\frac{Dw_{TL}}{Dt}\mathbf{\hat{e}_k}=-\bm{\Omega}_{flow}\times\mathbf{v}-2\bm{\Omega}\times\mathbf{v}-\frac{1}{\rho}\nabla p-g\mathbf{\hat{e}_k}+\mathbf{F},
\end{equation}
where $u_{TL}$, $v_{TL}$, and $w_{TL}$ are the tidally locked zonal, meridional, and vertical winds, respectively. $\mathbf{\hat{e}_i}$, $\mathbf{\hat{e}_j}$, and $\mathbf{\hat{e}_k}$ are the unit vectors of the tidally locked axes, respectively. $\bm{\Omega}_{flow}$ is the effective rotation rate \citep[\textcolor{black}{Equation 2.31 in}][]{Vallis_2019}, with formula of
\begin{equation}\label{eq:omega_flow}
    \textcolor{black}{\bm{\Omega}_{flow}=-\frac{v_{TL}}{r}\mathbf{\hat{e}_i}+\frac{u_{TL}}{r}\mathbf{\hat{e}_j}+\frac{u_{TL}\tan\phi_{TL}}{r}\mathbf{\hat{e}_k}},
\end{equation}
$r$ is the radial distance from the center of the planet, and \textcolor{black}{$\phi_{TL}$} is the tidally locked latitude. From Figure~\ref{fig:changing_coordinates}(b), $\bm{\Omega}$ can be written as 
\begin{equation}\label{eq:omega_frame}    
\textcolor{black}{\bm{\Omega}=\mit\Omega\sin\lambda_{TL}\mathbf{\hat{e}_i}+\mit\Omega\cos\lambda_{TL}\sin\phi_{TL}\mathbf{\hat{e}_j}-\mit\Omega\cos\lambda_{TL}\cos\phi_{TL}\mathbf{\hat{e}_k}},
\end{equation}
where $\mit\Omega$ represents the magnitude of planetary rotation rate, and \textcolor{black}{$\lambda_{TL}$} is the tidally locked longitude. Combining Equations~(\ref{eq:scalarform_momentum_equation})--(\ref{eq:omega_frame}) yields
\begin{equation}\label{dudt_TL}
    \textcolor{black}{\frac{Du_{TL}}{Dt}=-\frac{u_{TL}w_{TL}}{r}+\frac{\tan\phi_{TL}}{r}u_{TL}v_{TL}-\frac{1}{\rho r\cos\phi_{TL}}\frac{\partial p}{\partial\lambda_{TL}}-2\mit\Omega v_{TL}\cos\lambda_{TL}\cos\phi_{TL}-\rm 2\mit\Omega w_{TL}\cos\lambda_{TL}\sin\phi_{TL}+F_{\lambda}},
\end{equation}
\begin{equation}\label{dvdt_TL}
    \textcolor{black}{\frac{Dv_{TL}}{Dt}=-\frac{v_{TL}w_{TL}}{r}-\frac{\tan\phi_{TL}}{r}u_{TL}^2-\frac{1}{\rho r}\frac{\partial p}{\partial\phi_{TL}}+2\mit\Omega u_{TL}\cos\lambda_{TL}\cos\phi_{TL}+\rm 2\mit\Omega w_{TL}\sin\lambda_{TL}+F_{\phi}},
\end{equation}
\begin{equation}\label{dwdt_TL}
    \textcolor{black}{\frac{Dw_{TL}}{Dt}=\frac{u_{TL}^2+v_{TL}^2}{r}-\frac{1}{\rho}\frac{\partial p}{\partial r}-g+2\mit\Omega u_{TL}\cos\lambda_{TL}\sin\phi_{TL}-\rm 2\mit\Omega v_{TL}\sin\lambda_{TL}+F_r},
\end{equation}
where $F_{\lambda}$, $F_{\phi}$, and $F_{r}$ are the projections of the friction force on axes. On a typical terrestrial planet, the thickness of the atmosphere are usually ignored compared to its horizontal scale, and so is the vertical motion ($\omega$). Thus, the shallow atmosphere approximation ($r=a+z\approx a$, $a$ is planetary radius, and $z$ is height above surface; $\partial/\partial r\approx\partial/\partial z$; \textcolor{black}{$\left|w_{TL}\right|\ll\left|u_{TL}\right|,\left|v_{TL}\right|$}) is employed to simplify Equations~(\ref{dudt_TL})--(\ref{dwdt_TL}), shown as
\begin{equation}\label{simple_dudt_TL}
    \textcolor{black}{\frac{Du_{TL}}{Dt}=f_{TL}v_{TL}+\frac{\tan\phi_{TL}}{a}u_{TL}v_{TL}-\frac{1}{\rho a\cos\phi_{TL}}\frac{\partial p}{\partial\lambda_{TL}}+F_{\lambda}},
\end{equation}
\begin{equation}\label{simple_dvdt_TL}
    \textcolor{black}{\frac{Dv_{TL}}{Dt}=-f_{TL}u_{TL}-\frac{\tan\phi_{TL}}{a}u_{TL}^2-\frac{1}{\rho a}\frac{\partial p}{\partial\phi_{TL}}+F_{\phi}},
\end{equation}
\begin{equation}\label{simple_dwdt_TL}
    \frac{\partial p}{\partial z}=-\rho g,
\end{equation}
where \textcolor{black}{$f_{TL}\equiv-2\mit\Omega\cos\lambda_{TL}\cos\phi_{TL}$} is the Coriolis parameter in the tidally locked coordinates. The three momentum equations have the same forms as those in the standard spherical coordinates \citep[Equation 2.41 in][]{Vallis_2019}. Furthermore, Equations~(\ref{simple_dudt_TL})--(\ref{simple_dwdt_TL}) along with the unchanged Equations~(\ref{eq:continuity}) and (\ref{eq:thermodynamic}) suggest that the projected forms of the primitive equations in the two coordinates are the same except that the formulas of the Coriolis parameter and the velocities are different. Therefore, the processes to obtain the LEC in the standard coordinates are valid in the tidally locked coordinates.      

\begin{table}[]
    \centering
    \caption{General descriptions of symbols used in the LEC. Note that zonal mean is defined as the average along the standard longitudes in the standard coordinates, but along the tidally locked longitudes in the tidally locked coordinates. So does the deviation from the zonal mean.}
    \begin{tabular}{l l}
    \hline
    Symbols & Description \\
    \hline
    $X$ & Arbitrary quantity \\
    $\bar{X}$ & Temporal-mean of $X$ \\
    $\left[X\right]$ & Zonal-mean of $X$ \\
    $\tilde{X}$ & Global-mean of $X$ \\
    $X^{'}$ & Deviation from the temporal-mean of $X$, equal to $X-\bar{X}$ \\
    $X^{*}$ & Deviation from the zonal-mean of $X$, equal to $X-\left[X\right]$ \\
    $X^{''}$ & Deviation from the global-mean of $X$, equal to $X-\tilde{X}$ \\
    $C\left(X_1, X_2\right)$ & Conversion rate from $X_1$ to $X_2$ \\
    $G\left(X\right)$ & Generation rate of $X$ \\
    $D\left(X\right)$ & Dissipation rate of $X$ \\
    $\lambda$ & Longitude \\
    $\phi$ & Latitude \\
    $Z$ & Geopotential height \\
    $u$ & Zonal wind \\
    $v$ & Meridional wind \\
    $\omega$ & Vertical pressure velocity \\
    $a$ & Planetary radius (solid part) \\
    $g$ & Gravity \\
    $T$ & Air temperature \\
    $\theta$ & Air potential temperature \\
    $R$ & Gas constant for dry air \\
    $c_p$ & Specific heat at constant pressure \\
    $\kappa$ & $R/c_p$ \\
    $\gamma$ & Stability factor, equal to $-\frac{\kappa\theta}{pT}\left(\frac{\partial\tilde{\theta}}{\partial p}\right)^{-1}$ \\
    $dm$ & Mass element, equal to $a^2\cos\phi d\lambda d\phi dp/g$ \\
    \hline
    \end{tabular}
    \label{tab:my_label_1}
\end{table}

Following section 14.3 in \cite{Peixoto_1992}, we combine Equations~(\ref{eq:continuity}), (\ref{eq:thermodynamic}), and (\ref{simple_dudt_TL})--(\ref{simple_dwdt_TL}), and then integrate over the whole atmosphere to disregard all boundary terms. This process yields
\begin{equation}\label{dPM/dt}
    \frac{\partial \rm P_M}{\partial t}=-{\rm C\left(P_M, P_E\right)}-{\rm C\left(P_M, K_M\right)}+\rm {G\left(P_M\right)},
\end{equation}
\begin{equation}\label{dPE/dt}
    \frac{\partial \rm P_E}{\partial t}={\rm C\left(P_M, P_E\right)}-{\rm C\left(P_E, K_E\right)}+\rm {G\left(P_E\right)},
\end{equation}
\begin{equation}\label{dKM/dt}
    \frac{\partial \rm K_M}{\partial t}={\rm C\left(K_E, K_M\right)}+{\rm C\left(P_M, K_M\right)}-\rm {D\left(K_M\right)},
\end{equation}
\begin{equation}\label{dKE/dt}
    \frac{\partial \rm K_E}{\partial t}=-{\rm C\left(K_E, K_M\right)}+{\rm C\left(P_E, K_E\right)}-\rm {D\left(K_E\right)},
\end{equation}
where P$_{\rm M}$ is mean potential energy, P$_{\rm E}$ is eddy potential energy, K$_{\rm M}$ is mean kinetic energy, K$_{\rm E}$ is eddy kinetic energy, $\rm C\left(X_1, X_2\right)$ is conversion rate from X$_1$ to X$_2$, $\rm G\left(X\right)$ is generation rate of X, and $\rm D\left(X\right)$ is dissipation rate of X (Figure~\ref{fig:LEC_framework}). The general descriptions of symbols we used are shown in Table~\ref{tab:my_label_1}. The detailed descriptions of these terms are shown in Appendix~\ref{appendix_b}.

\subsection{Data}
Daily-mean data for the atmosphere on Earth is from the National Center for Environmental Prediction and the Department of Energy reanalysis datasets (NCEP R2). These datasets are produced by an advanced data assimilation method combining numerical models and observational data. The spatial resolution is $2.5^{\circ}\times2.5^{\circ}$ in latitude and longitude with 17 levels from surface to 10 hPa. The period covered is the whole year of 1979. The data is grouped by 12 months and the LEC is calculated in each month and then averaged to get the annual-mean LEC. This evaluation is based on the fact that there are strong seasonal cycles and that synoptic eddies have lifetimes of mostly several days and less than a month.

Daily-mean data for the atmosphere on tidally locked terrestrial planets is from simulations by the Exoplanet Community Atmosphere Model \citep[ExoCAM,][]{Wolf_2014,Wolf_2015,Wolf_et_al_2017,Wolf_2022}. One simulation is for a rapidly rotating, tidally locked planet with a rotation period (= orbital period) of 5 Earth days, and the other simulation is for a slowly rotating, tidally locked planet with a rotation period (= orbital period) of 60 Earth days. In the two simulations, \textcolor{black}{the solar constant is 1360 W\,m$^{-2}$, planets are Earth-sized aquaplanets with terrestrial atmospheres (1 bar N$_2$, 400 ppmv CO$_2$, and flexible water vapor) and the same gravity as Earth}. The spatial resolution is $4^{\circ}\times5^{\circ}$ in latitude and longitude with 40 pressure levels from surface to 10 Pa. The period covered is the last 300 model days. We directly calculate the LEC over all 300 days instead of month by month, based on that there is no seasonal or diurnal cycle on the 1:1 tidally locked planets.

In our calculations, all GCMs' data over 100 hPa are excluded. This is because temperatures and winds are strongly model-dependent near the top of the atmosphere \citep{Sergeev_2022,Turbet_2022}. For consistency, the NCEP R2 data over 100 hPa are also excluded. These treatments do not affect the main conclusions in this work.

\section{Results}\label{sec:result}
\subsection{Thermal structure and atmospheric circulation}\label{structure_and_circulation}
\begin{figure*}
    \centering
    \includegraphics[width=0.9\textwidth]{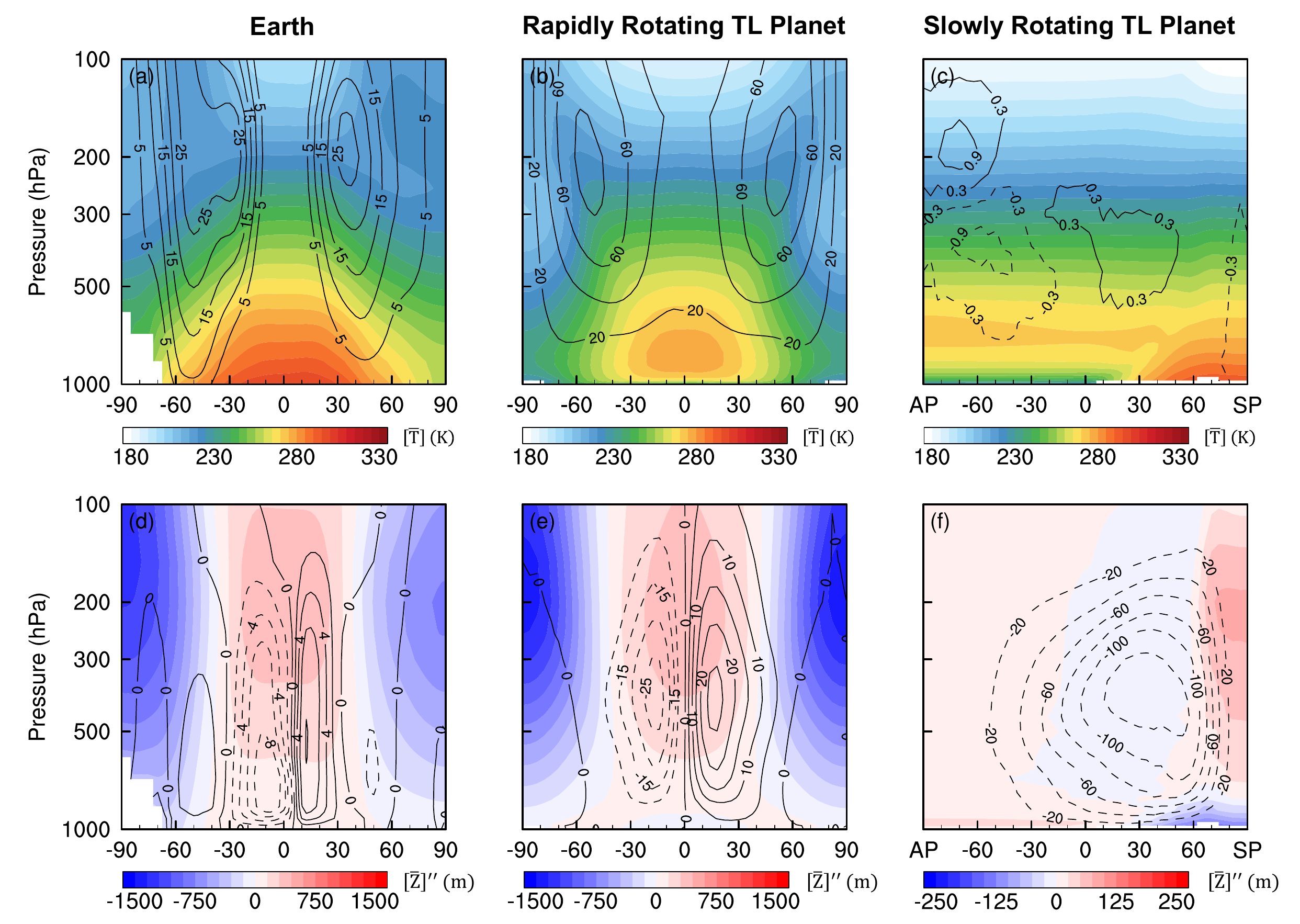}
    \caption{Climatic elements on three planets. (a)--(c) Temporal- and zonal-mean temperature (color shading) and zonal-mean zonal winds (contour lines) on Earth, on the rapidly rotating tidally locked terrestrial planet, and on the slowly rotating tidally locked terrestrial planet; (d)--(f) same as (a)--(c) but for deviations of temporal- and zonal-mean geopotential height from the global means (color shading), and mass stream functions (contour lines, in units of $10^{10}$ kg\,s$^{-1}$). Note that the ranges of color bars in panels (d), (e), and (f) are different. Panels (a), (b), (d) and (e) are shown in the standard coordinates, and (c) and (f) are shown in the tidally locked coordinates. SP and AP are the substellar point and the antistellar point, respectively.}
    \label{fig:climate_three_planets}
\end{figure*}

\begin{figure}
    \centering
    \includegraphics[width=0.45\textwidth]{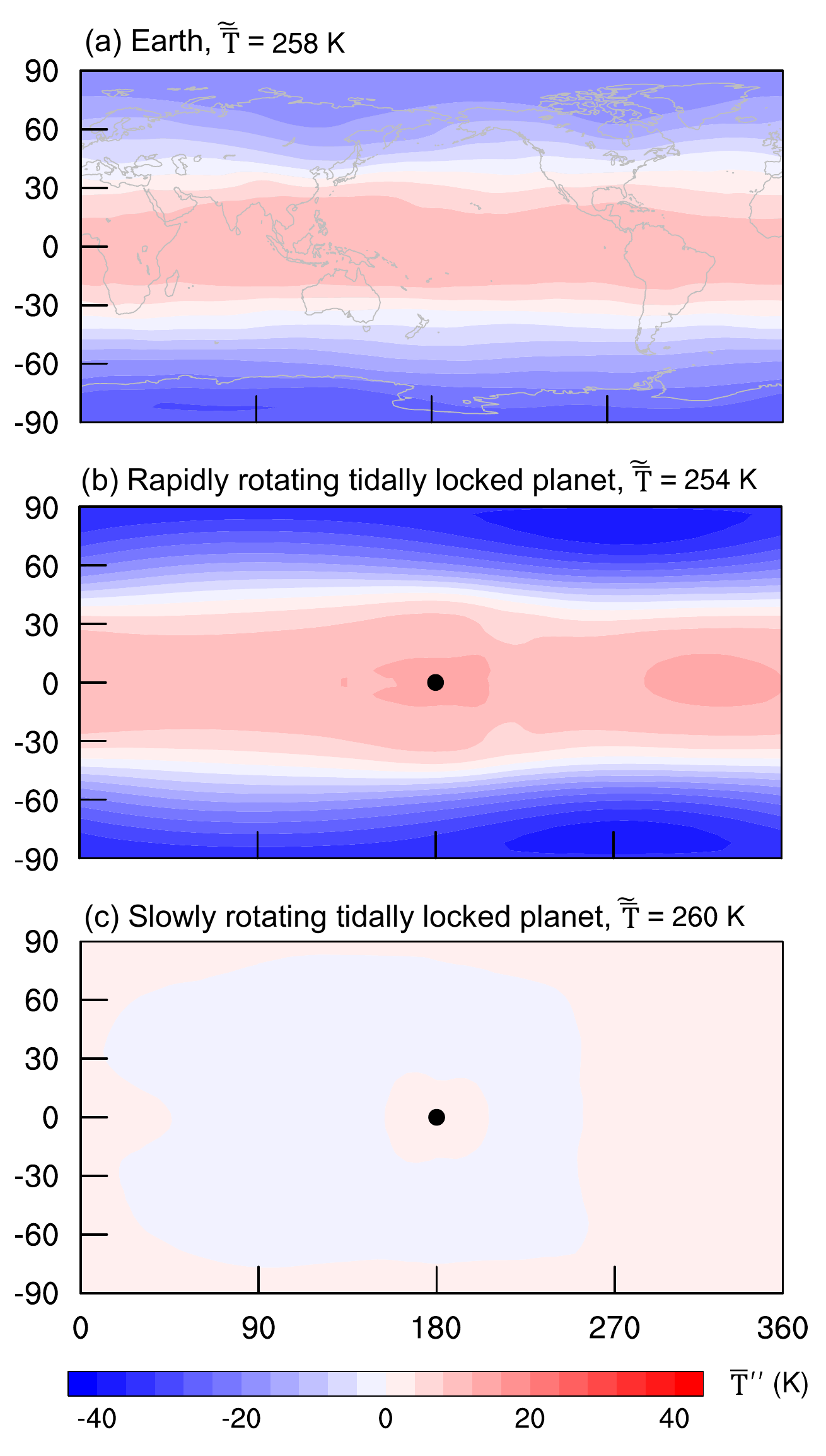}
    \caption{Deviations of temporal-mean temperature from the global means at 500 hPa on (a) Earth, (b) one rapidly rotating tidally locked terrestrial planet, and (c) one slowly rotating tidally locked terrestrial planet. The temporal- and global-mean temperatures are shown in the top of each panel. The black dots represent substellar points.}
    \label{fig:temp500_three_planets}
\end{figure}

\textcolor{black}{For Earth, the rapidly rotating tidally locked planet, and the slowly rotating tidally locked planet, the solar constant is 1360 W\,m$^{-2}$. It corresponds to a global mean of 340 W\,m$^{-2}$ received stellar radiation at the top of the atmosphere, but the absorbed values are different for the three planets, suggesting different efficiencies of energy input to the planets. On Earth, the absorbed stellar radiation is 238 W\,m$^{-2}$, including 80 W\,m$^{-2}$ absorbed by the atmosphere and 158 W\,m$^{-2}$ absorbed by the surface. The global-mean surface temperature is 288 K. The absorbed stellar radiation on the tidally locked planets is smaller than that on Earth, i.e. 187 W\,m$^{-2}$ on the rapidly rotating planet (57 W\,m$^{-2}$ absorbed by the atmosphere and 130 W\,m$^{-2}$ absorbed by the surface) and 175 W\,m$^{-2}$ on the slowly rotating planet (57 W\,m$^{-2}$ absorbed by the atmosphere and 118 W\,m$^{-2}$ absorbed by the surface), due to the reflection by thick clouds near the substellar location \citep{Yang_2013}. Naturally, the global-mean surface temperatures on the two tidally locked planets are lower than that on Earth, i.e. 257 K on the rapidly rotating planet and 247 K on the slowly rotating planet.}

The \textcolor{black}{air} temperature, geopotential height, and atmospheric circulation on \textcolor{black}{the three planets} are shown in Figure~\ref{fig:climate_three_planets}. In the free atmosphere, these climatic elements are zonally homogeneous on the rapidly rotating tidally locked terrestrial planet and axisymmetric on the slowly rotating tidally locked terrestrial planet, so that the former is shown in the standard coordinates and the latter is shown in the tidally locked coordinates.   

The temperature structure on Earth and on the rapidly rotating tidally locked terrestrial planet are analogous, as they show steep meridional (south--north) temperature gradients in the middle latitudes (Figures~\ref{fig:climate_three_planets}(a) and (b)). While on the slowly rotating tidally locked terrestrial planet, the global free atmosphere is in a weak temperature gradient (WTG) regime \citep{Pierrehumbert_2010,Pierrehumbert_Hammond_2019}, for which the air temperature is almost horizontally homogeneous everywhere (Figure~\ref{fig:climate_three_planets}(c)). This is because weak Coriolis effect cannot maintain a large pressure or temperature gradient. The WTGs can be more obviously seen in the horizontal temperature structures. For example, the equator--pole temperature difference at 500 hPa is about 40 K on Earth and about 60 K on the rapidly rotating tidally locked terrestrial planet (Figures~\ref{fig:temp500_three_planets}(a) and (b)). However, the temperature difference on the slowly rotating tidally locked terrestrial planet is no more than 5 K (Figure~\ref{fig:temp500_three_planets}(c)). The WTGs are destroyed only very close to the surface, due to the effect of surface friction. Note that in the WTG regime, the difference of geopotential height is also small, e.g., no more than 500 m on the slowly rotating planet, while this value is about 3000 m on Earth and on the rapidly rotating planet (Figures~\ref{fig:climate_three_planets}(d)--(f)). 

Figure~\ref{fig:climate_three_planets}(c) shows a strong temperature inversion on the slowly rotating tidally locked planet, which is mainly on the nightside and extends to the dayside. It is caused by the uneven distribution in the stellar radiation and the effective energy transport from dayside to nightside in the free atmosphere \textcolor{black}{\citep{Joshi_2020}}. The strong inversion can make the atmosphere be stable and inhibit the growth of eddies, which can influence the LEC and will be shown in the subsequent sections.

On Earth, the annual- and zonal-mean mass stream functions clearly show the Hadley cells in the tropics and the Ferrel cells in the middle latitudes (Figure~\ref{fig:climate_three_planets}(d)). On the rapidly rotating tidally locked terrestrial planet, the Hadley cells expand and become dominant while the Ferrel cells almost disappear (Figure~\ref{fig:climate_three_planets}(e)). This is due to that the planetary rotation rate is 1/5 of that on Earth. The strength of Hadley cells on this rapidly rotating planet is about $25\times10^{10}$ kg\,s$^{-1}$, and is somewhat larger than the value of $10\times10^{10}$ kg\,s$^{-1}$ on Earth. Differently, the circulation on the slowly rotating tidally locked terrestrial planet is an isotropic overturning circulation around the substellar point (Figure~\ref{fig:climate_three_planets}(f)). It extends to the globe due to the very small rotation rate, which is 1/60 of that on Earth. The strength of the global overturning circulation is about $120\times10^{10}$ kg\,s$^{-1}$, and is much stronger than the Hadley cells on Earth and on the rapidly rotating planet. This is due to that the day--night surface temperature contrast is about 100 K, much larger than the equator--pole surface temperature difference on Earth, about 50 K (Figures~\ref{fig:changing_coordinates}(c) and (d)).   

\subsection{LEC on Earth}\label{Earth}
\begin{figure}
    \centering
    \includegraphics[width=0.45\textwidth]{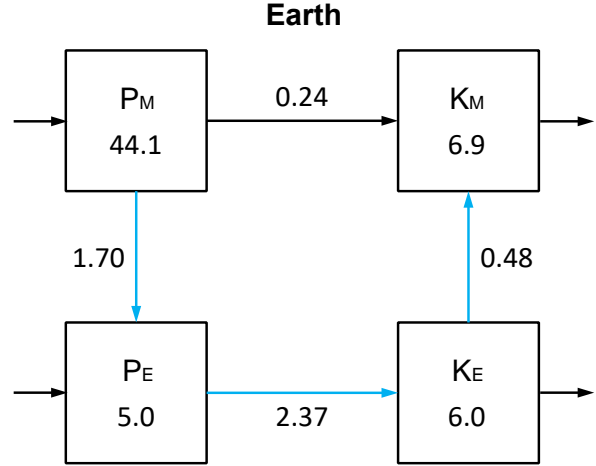}
    \caption{The global-mean vertical integrals of energy components (in boxes) and conversion rates (close to arrows) on Earth. Units are $10^5$ J\,m$^{-2}$ for energy and W\,m$^{-2}$ for conversion rates. \textcolor{black}{Arrows without specific values represent generation rates and dissipation rates, same as those in Figure~\ref{fig:LEC_framework}(a).} \textcolor{black}{Directions of arrows} represent the directions of energy conversion, and the main conversion paths are highlighted by blue color. Values are calculated in the standard coordinates.}
    \label{fig:LEC_diagram_1}
\end{figure}

The global-mean vertical integrals of the LEC on Earth are shown in Figure~\ref{fig:LEC_diagram_1}. The total available energy stored in the atmosphere is about $62.0\times10^5$ J\,m$^{-2}$, including $44.1\times10^5$ J\,m$^{-2}$ of P$_{\rm M}$, $5.0\times10^5$ J\,m$^{-2}$ of P$_{\rm E}$, $6.9\times10^5$ J\,m$^{-2}$ of K$_{\rm M}$, and $6.0\times10^5$ J\,m$^{-2}$ of K$_{\rm E}$. The main conversion path is P$_{\rm M}\rightarrow$P$_{\rm E}\rightarrow$K$_{\rm E}\rightarrow$K$_{\rm M}$. P$_{\rm M}$ is converted to P$_{\rm E}$ at a rate of 1.70 W\,m$^{-2}$ through heat transport by baroclinic eddies. P$_{\rm E}$ is converted to K$_{\rm E}$ at a rate of 2.37 W\,m$^{-2}$ through cross-isobaric motions in baroclinic eddies. Some portion of K$_{\rm E}$ is converted to K$_{\rm M}$ at a rate of 0.48 W\,m$^{-2}$ through wave--mean flow interactions. In addition, P$_{\rm M}$ is ultimately converted to K$_{\rm M}$ at a relatively inefficient rate of 0.24 W\,m$^{-2}$ through cross-isobaric motions in the Hadley and Ferrel cells. Our results are consistent with previous estimations \citep[e.g.,][]{Peixoto_1974,Li_2007,Kim_2013}. We also recalculate the LEC using the data simulated by ExoCAM, and obtain analogous results to those based on reanalysis data (figures not shown).   

\textcolor{black}{Note that the four arrows in Figure~\ref{fig:LEC_diagram_1} without specific values represent the generation rates of P$_{\rm M}$ and P$_{\rm E}$ and the dissipation rates of K$_{\rm M}$ and K$_{\rm E}$, respectively. It is difficult to calculate directly from the reanalysis data, so we omit their values here. However, their values can be estimated by assuming an equilibrium energy cycle. That is, P$_{\rm M}$ is converted to P$_{\rm E}$ and K$_{\rm M}$ with 1.70 and 0.24 W\,m$^{-2}$, respectively, and therefore a generation rate of P$_{\rm M}$ with 1.94 W\,m$^{-2}$ is required. Similarly, the generation rate of P$_{\rm E}$ is 0.67 W\,m$^{-2}$, the dissipation rate of K$_{\rm M}$ is 0.72 W\,m$^{-2}$, and the dissipation rate of K$_{\rm E}$ is 1.89 W\,m$^{-2}$. We omit the corresponding results on the other two planets, and discuss them in Summary and Discussions.}

\begin{figure*}
    \centering
    \includegraphics[width=0.9\textwidth]{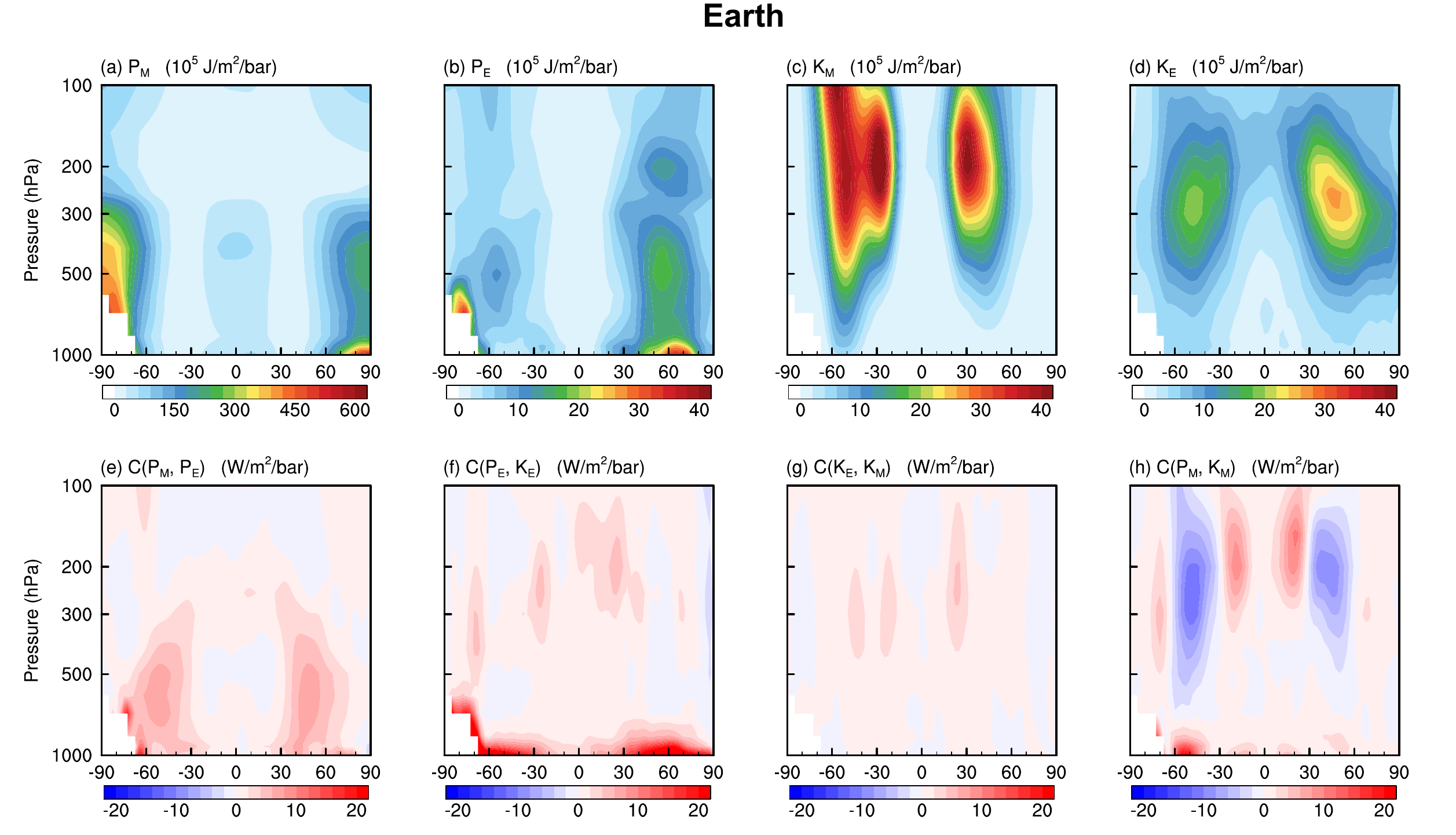}
    \caption{Latitude-altitude cross-sections of the LEC on Earth. (a)--(d) Mean potential energy (P$_{\rm M}$), eddy potential energy (P$_{\rm E}$), mean kinetic energy (K$_{\rm M}$), and eddy kinetic energy (K$_{\rm E}$), respectively; (e)--(h) conversion rates of mean potential energy to eddy potential energy, eddy potential energy to eddy kinetic energy, eddy kinetic energy to mean kinetic energy, and mean potential energy to mean kinetic energy, respectively. Global-mean vertical integrals of each panel are corresponding values in Figure~\ref{fig:LEC_diagram_1}. Panels are shown in the standard coordinates. The range of color bar in panel (a) is different from it in panels (b)--(d).}
    \label{fig:LEC_Earth}
\end{figure*}

The majority of P$_{\rm M}$ lies in the range of 1000--300 hPa over the polar regions, especially over the Antarctic continent, where the temperature departure from the global mean is largest (Figures \ref{fig:climate_three_planets}(a) and \ref{fig:LEC_Earth}(a)). There is also a secondary maximum near the tropopause in the tropics, as the temperature there also deviates significantly from the global mean, although the meridional temperature gradient is weak. P$_{\rm E}$ is mainly in 30--90$^{\circ}$S/N (Figure~\ref{fig:LEC_Earth}(b)), resulting from temperature anomalies induced by baroclinic eddies, forced orographic waves, and land--sea surface temperature contrast \citep{Li_2007}. Most of K$_{\rm M}$ is in the troposphere over the middle latitudes of 30--60$^{\circ}$S/N, as it reflects the subtropical and mid-latitude jets (Figures~\ref{fig:climate_three_planets}(a) and \ref{fig:LEC_Earth}(c)). In addition, there is an extra maximum extending into the stratosphere in the southern hemisphere, as it reflects the stratospheric jet (over 100 hPa, not entirely shown). K$_{\rm E}$ is mainly centered in the troposphere over 50--60$^{\circ}$S/N (Figure~\ref{fig:LEC_Earth}(d)), associated with the storm tracks over the Pacific, Atlantic, and Southern oceans \citep{Trenberth_1991,Harnik_2003}.

The structure of P$_{\rm E}$ is analogous to that of P$_{\rm M}$ except for a slightly shift to the equator, \textcolor{black}{since P$_{\rm E}$ is conversed from P$_{\rm M}$ through baroclinic eddies so that is} affected by P$_{\rm M}$. Likewise, the structure of K$_{\rm E}$ is analogous to that of K$_{\rm M}$. Besides, K$_{\rm E}$ \textcolor{black}{is also affected by P$_{\rm E}$ as it is conversed from P$_{\rm E}$}. Thus, the maxima of K$_{\rm E}$ are between the maxima of P$_{\rm E}$ and K$_{\rm M}$, suggesting K$_{\rm E}$ is a result of a balance between P$\rm_{E}$ and K$_{\rm M}$.

The conversion from P$_{\rm M}$ to P$_{\rm E}$ occurs mainly in the middle troposphere over the middle latitudes of 30--60$^{\circ}$S/N, where the baroclinic eddies are strong (Figure~\ref{fig:LEC_Earth}(e)). The growing baroclinic eddies transport heat poleward and reduce the south--north temperature gradients, but lead additional east--west temperature variance, i.e., reduce P$_{\rm M}$ but generate P$_{\rm E}$. Inside these eddies, the cross-isobaric motion converts a small portion of P$_{\rm E}$ to K$_{\rm E}$ (Figure~\ref{fig:LEC_Earth}(f)). P$_{\rm E}$ is converted to K$_{\rm E}$ mainly near the surface over mid-to-high latitudes, especially over the Antarctic, and is associated with the heat-driven rising and sinking motions \citep{Li_2007}. In the middle troposphere over the middle latitudes, eddies are generate and then propagate out of this region, but bring the momentum back to accelerate the jet, and these wave--mean flow interactions convert K$_{\rm E}$ to K$_{\rm M}$ (Figure~\ref{fig:LEC_Earth}(g)). P$_{\rm M}$ is converted to K$_{\rm M}$ in the tropics and subtropics by the motion along the pressure gradient in the Hadley cells nearly following angular momentum conservation, but K$_{\rm M}$ is converted back to P$_{\rm M}$ in the middle latitudes by the motion against the pressure gradient in the Ferrel cells (Figures~\ref{fig:climate_three_planets}(d) and \ref{fig:LEC_Earth}(h)). The combined action of the conversion between P$_{\rm M}$ and K$_{\rm M}$ makes the global-mean conversion from P$_{\rm M}$ to K$_{\rm M}$ be relatively inefficient, and even be negative sometimes \citep[e.g., Figure 14.8 in][]{Peixoto_1992}.

The LEC perspective is consistent with the momentum budget. The maxima of K$_{\rm M}$ over 40--60$^{\circ}$S/N coincide with the regions where K$_{\rm E}$ is converted to K$_{\rm M}$ but K$_{\rm M}$ is converted to P$_{\rm M}$. It indicates that the mid-latitude jets are eddy-driven---the poleward momentum transport by eddies maintains the mid-latitude jets. Likewise, the maxima of K$_{\rm M}$ in the subtropics coincide with the regions where both P$_{\rm M}$ and K$_{\rm E}$ are converted to K$_{\rm M}$, indicating that the poleward angular momentum transport by meridional circulation and eddies together maintain the subtropical jets.

\subsection{LEC on rapidly rotating, tidally locked terrestrial planet}\label{rapid}
\begin{figure*}
    \centering
    \includegraphics[width=0.9\textwidth]{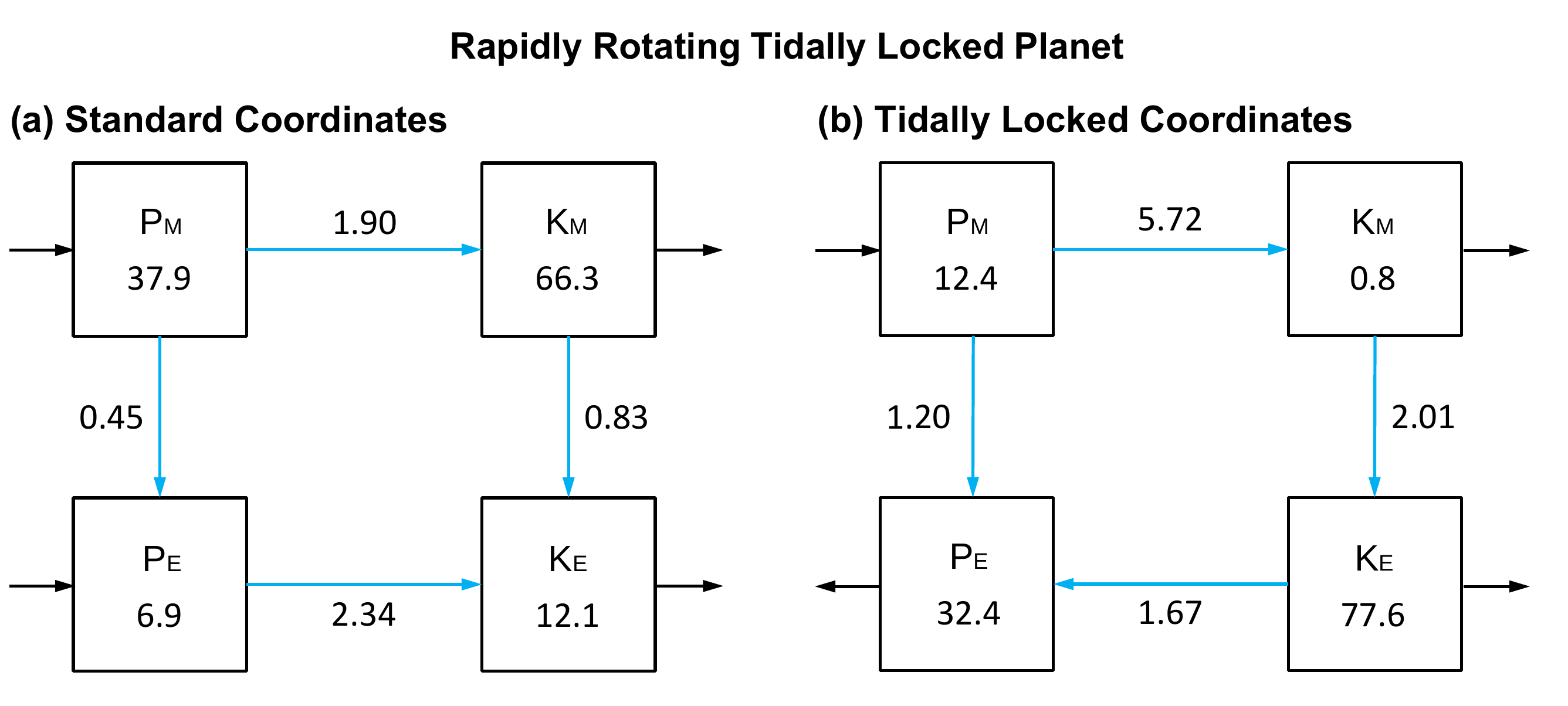}
    \caption{Same as Figure~\ref{fig:LEC_diagram_1} but on the rapidly rotating tidally locked terrestrial planet. Values are calculated in the standard coordinates (a) and in the tidally locked coordinates (b).}
    \label{fig:LEC_diagram_3}
\end{figure*}

Figure~\ref{fig:LEC_diagram_3}(a) displays the global-mean vertical integrals of the LEC on the rapidly rotating tidally locked terrestrial planet. In our estimations, P$_{\rm M}$ is about $37.9\times10^5$ J\,m$^{-2}$, P$_{\rm E}$ is about $6.9\times10^5$ J\,m$^{-2}$, K$_{\rm M}$ is about $66.3\times10^5$ J\,m$^{-2}$, and K$_{\rm E}$ is about $12.1\times10^5$ J\,m$^{-2}$. There are two main conversion paths, P$_{\rm M}\rightarrow$P$_{\rm E}\rightarrow$K$_{\rm E}$ and P$_{\rm M}\rightarrow$K$_{\rm M}\rightarrow$K$_{\rm E}$. On the one hand, P$_{\rm M}$ is converted to P$_{\rm E}$ at a rate of 0.45 W\,m$^{-2}$, and then into K$_{\rm E}$ at a rate of 2.34 W\,m$^{-2}$. On the other hand, P$_{\rm M}$ is converted to K$_{\rm M}$ at a rate of 1.90 W\,m$^{-2}$, and then into K$_{\rm E}$ at a rate of 0.83 W\,m$^{-2}$.        

\begin{figure*}
    \centering
    \includegraphics[width=0.9\textwidth]{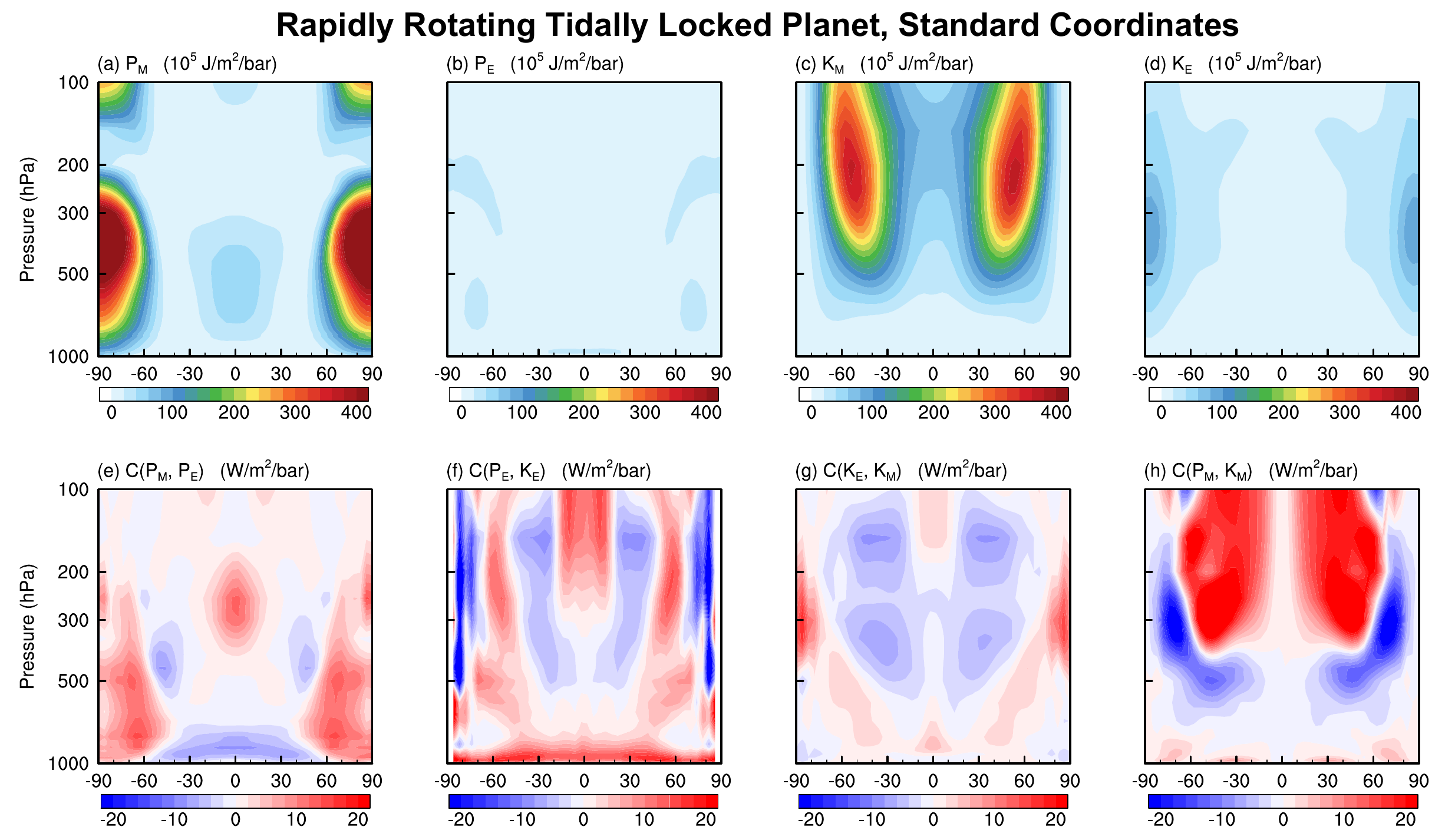}
    \caption{Same as Figure~\ref{fig:LEC_Earth} but on the rapidly rotating tidally locked terrestrial planet. Panels are shown in the standard coordinates. The maximum in panel (a) is about $600\times10^5$ J\,m$^{-2}$\,bar$^{-1}$; the maximum in panel (h) is about $40$ W\,m$^{-2}$\,bar$^{-1}$, and the minimum is about $-30$ W\,m$^{-2}$\,bar$^{-1}$.}
    \label{fig:LEC_Fast_TL_ST}
\end{figure*}

\textcolor{black}{The stellar flux received by the planet has an equator--pole contrast, which can generate P$_{\rm M}$ in the atmosphere.} The majority of P$_{\rm M}$ is in the middle troposphere over the polar regions, and a portion is in the tropics (Figure~\ref{fig:LEC_Fast_TL_ST}(a)). The structure is similar to that on Earth, because they both have similar south--north temperature gradient (Figures~\ref{fig:climate_three_planets}(a), \ref{fig:climate_three_planets}(b), \ref{fig:temp500_three_planets}(a), and \ref{fig:temp500_three_planets}(b)). \textcolor{black}{P$_{\rm E}$ is much smaller than P$_{\rm M}$ and mostly over the polar regions, with contributions from the extratropical stationary Rossby waves (Figure~\ref{fig:LEC_Fast_TL_ST}(b)).} The cold lobes of the Rossby waves are visible in the horizontal temperature structure (Figure~\ref{fig:temp500_three_planets}(b)). K$_{\rm M}$ is located in the mid-to-upper troposphere over the middle latitudes, coinciding with the two westerly jets (Figures~\ref{fig:climate_three_planets}(b) and \ref{fig:LEC_Fast_TL_ST}(c)). Moreover, K$_{\rm M}$ over the equator is also non-negligible due to the equatorial superrotation. K$_{\rm E}$ is concentrated in the polar regions, associated with the extratropical stationary Rossby waves (Figure~\ref{fig:LEC_Fast_TL_ST}(d)).

On the rapidly rotating tidally locked planets, although the received stellar radiation and the underlying surface are invariant, the atmosphere still has large variability, i.e., transient eddies, which can be seen in the instantaneous pressure field, temperature, winds, water vapor, and clouds \citep[e.g.,][]{Merlis_2010,Pierrehumbert_Hammond_2019,Song_2021}. The variability also contributes a portion of P$_{\rm E}$ and K$_{\rm E}$.

\begin{figure*}
    \centering
    \includegraphics[width=0.9\textwidth]{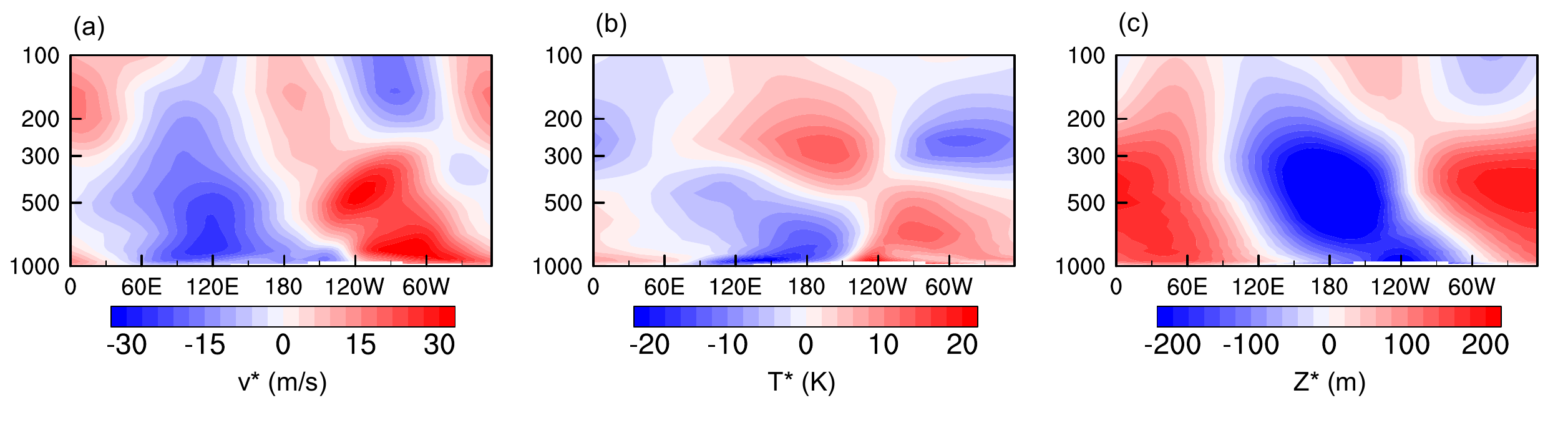}
    \caption{Longitude-altitude cross-sections of (a) eddy meridional velocity, (b) eddy temperature, and (c) eddy geopotential height at 70$^{\circ}$N on the rapidly rotating tidally locked terrestrial planet. Panels are instantaneous quantities departed from their zonal means.}
    \label{fig:baroclinic eddies}
\end{figure*}

In the middle troposphere over the polar regions, \textcolor{black}{P$_{\rm M}$ is converted to P$_{\rm E}$ by eddies (Figure~\ref{fig:LEC_Fast_TL_ST}(e)). These eddies are conjectured to arise from a form of baroclinic instability \citep{Pierrehumbert_Hammond_2019}. Figure~\ref{fig:baroclinic eddies} shows the instantaneous longitude-altitude structure of these eddies with a slightly westward tilt, which is similar to the unstable baroclinic mode on Earth, but on a larger scale due to the smaller planetary rotation rate \citep{Holton_2013}. Meanwhile,} in the upper troposphere over the equator, the conversion from P$_{\rm M}$ to P$_{\rm E}$ is caused by the equatorial wave activity. However, a fraction of P$_{\rm E}$ is converted back to P$_{\rm M}$ in the range of 500--300 hPa over the middle latitudes and near the surface, which is due to the up-gradient heat transport by stationary waves. The cross-isobaric motion converts P$_{\rm E}$ to K$_{\rm E}$ in the tropics and middle latitudes, but converts K$_{\rm E}$ back to P$_{\rm E}$ in the subtropics and polar regions (Figure~\ref{fig:LEC_Fast_TL_ST}(f)). In general, the net conversion is from P$_{\rm E}$ to K$_{\rm E}$. In the free troposphere, K$_{\rm E}$ is converted to K$_{\rm M}$ over the tropics, but K$_{\rm M}$ is converted to K$_{\rm E}$ over the middle latitudes, which is caused by the wave--mean flow interactions (Figure~\ref{fig:LEC_Fast_TL_ST}(g)). That is, the stationary Rossby and Kelvin waves form as a Matsuno-Gill mode that transports westerly momentum from the middle latitudes to the equator, accelerating the equatorial jet but damping the jets in the middle latitudes \citep[e.g.,][]{Matsuno_1966,Gill_1980,Showman_2011}. However, the net conversion is from K$_{\rm M}$ to K$_{\rm E}$. The conversion from P$_{\rm M}$ to K$_{\rm M}$ is dominant from the tropics to the middle latitudes through the expanded Hadley cells, while the conversion from K$_{\rm M}$ back to P$_{\rm M}$ only occurs at high latitudes through the weak and narrow Ferrel cells (Figures~\ref{fig:climate_three_planets}(e) and \ref{fig:LEC_Fast_TL_ST}(h)). Thus, the net conversion is from P$_{\rm M}$ to K$_{\rm M}$.

The maxima of K$_{\rm M}$ over 30--60$^{\circ}$S/N coincide with the regions where P$_{\rm M}$ is converted to K$_{\rm M}$. It suggests that the jets there are maintained by poleward angular momentum transport through the Hadley cells rather than by momentum transport through waves. Zonal jets constrained by the angular momentum conservation would obey
\begin{equation}\label{eq:angular_momentum_conservation}
\left[u\right]=\Omega a\frac{\sin^2\phi}{\cos\phi},    
\end{equation}
where $\Omega$ is the planetary rotation rate, $a$ is the planetary radius, $\phi$ is the latitude, and wind speed over the equator is assumed to be zero \citep[\textcolor{black}{Equation 11.4 in}][]{Vallis_2019}. It gives an estimate of the zonal-mean wind speed at 50$^{\circ}$S/N to be about 80 m\,s$^{-1}$, in agreement with the simulation. Thus, the jets are more like subtropical jets, because their driving mechanism is similar to that of the subtropical jets on Earth.

Comparing to Earth, a main difference of the LEC is that K$_{\rm M}$ is much larger, i.e., about ten times larger, which is related to the larger wind speeds. For example, a zonal jet has a maximum wind speed of about 80 m\,s$^{-1}$ on the rapidly rotating tidally locked planet, while this value is only 30 m\,s$^{-1}$ on Earth (Figures~\ref{fig:climate_three_planets}(a) and (b)). This is due to that wider Hadley cells move the air further away from the equator, allowing larger wind speeds, although the planet rotates at 1/5 the rate of Earth (Equation~(\ref{eq:angular_momentum_conservation})). The stronger winds are also consistent with the larger meridional temperature gradient in the high latitudes of the planet (Figure~\ref{fig:temp500_three_planets}(b)). Another big difference from Earth is that the net conversion from P$_{\rm M}$ to K$_{\rm M}$ becomes efficient. This is due to that the Hadley cells become stronger and wider, but the Ferrel cells become weaker, as a result of the smaller rotation rate. This situation is similar to Venus, where the Hadley cells extend throughout the atmosphere and make the conversion from P$_{\rm M}$ to K$_{\rm M}$ be most efficient \citep{Lee_2010}. In addition, K$_{\rm M}$ is eventually converted to K$_{\rm E}$ rather than K$_{\rm E}$ being converted to K$_{\rm M}$.
      
\subsection{LEC on slowly rotating, tidally locked terrestrial planet}\label{slow}
\begin{figure*}
    \centering
    \includegraphics[width=0.9\textwidth]{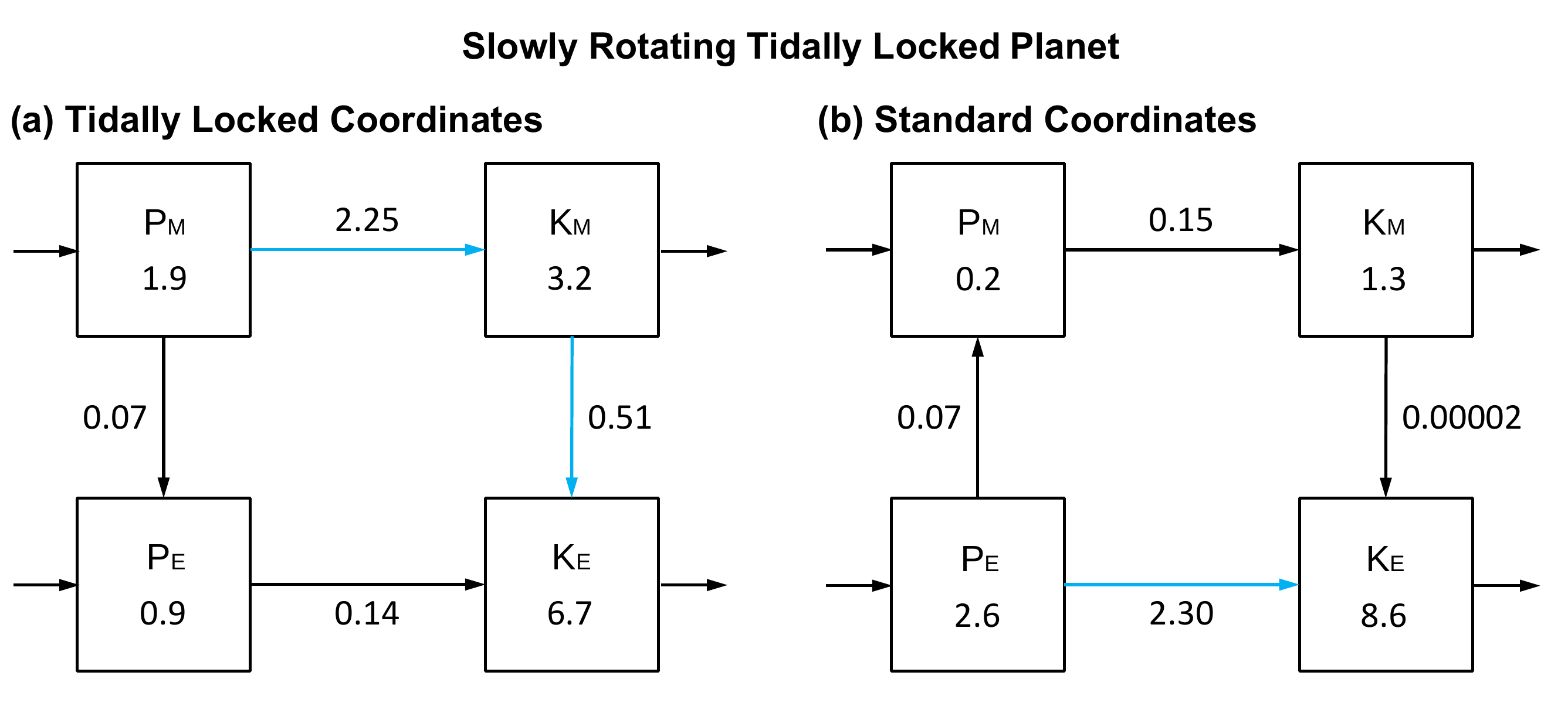}
    \caption{Same as Figure~\ref{fig:LEC_diagram_1} but on the slowly rotating tidally locked terrestrial planet. Values are calculated in the tidally locked coordinates (a) and in the standard coordinates (b).}
    \label{fig:LEC_diagram_2}
\end{figure*}

In this section, the LEC on the slowly rotating tidally locked planet is calculated using the tidally locked formulas. The global-mean vertical integrals are shown in Figure~\ref{fig:LEC_diagram_2}(a). Overall, the total available energy is about $12.7\times10^5$ J\,m$^{-2}$, including $1.9\times10^5$ J\,m$^{-2}$ of P$_{\rm M}$, $0.9\times10^5$ J\,m$^{-2}$ of P$_{\rm E}$, $3.2\times10^5$ J\,m$^{-2}$ of K$_{\rm M}$, and $6.7\times10^5$ J\,m$^{-2}$ of K$_{\rm E}$. The main conversion path is P$_{\rm M}\rightarrow$K$_{\rm M}\rightarrow$K$_{\rm E}$: P$_{\rm M}$ is converted to K$_{\rm M}$ at a rate of 2.25 W\,m$^{-2}$ through cross-isobaric motions in global overturning circulation, and a portion of K$_{\rm M}$ is converted to K$_{\rm E}$ at a rate of 0.51 W\,m$^{-2}$ through interactions between the overturning circulation and eddies. While the other path, P$_{\rm M}\rightarrow$P$_{\rm E}\rightarrow$K$_{\rm E}$, which occurs in baroclinic eddies and stationary planetary waves, is relatively inefficient, as that the conversion rate from P$_{\rm M}$ to P$_{\rm E}$ is only 0.07 W\,m$^{-2}$ and the conversion rate from P$_{\rm E}$ to K$_{\rm E}$ is only 0.14 W\,m$^{-2}$.    

\begin{figure*}
    \centering    
    \includegraphics[width=0.9\textwidth]{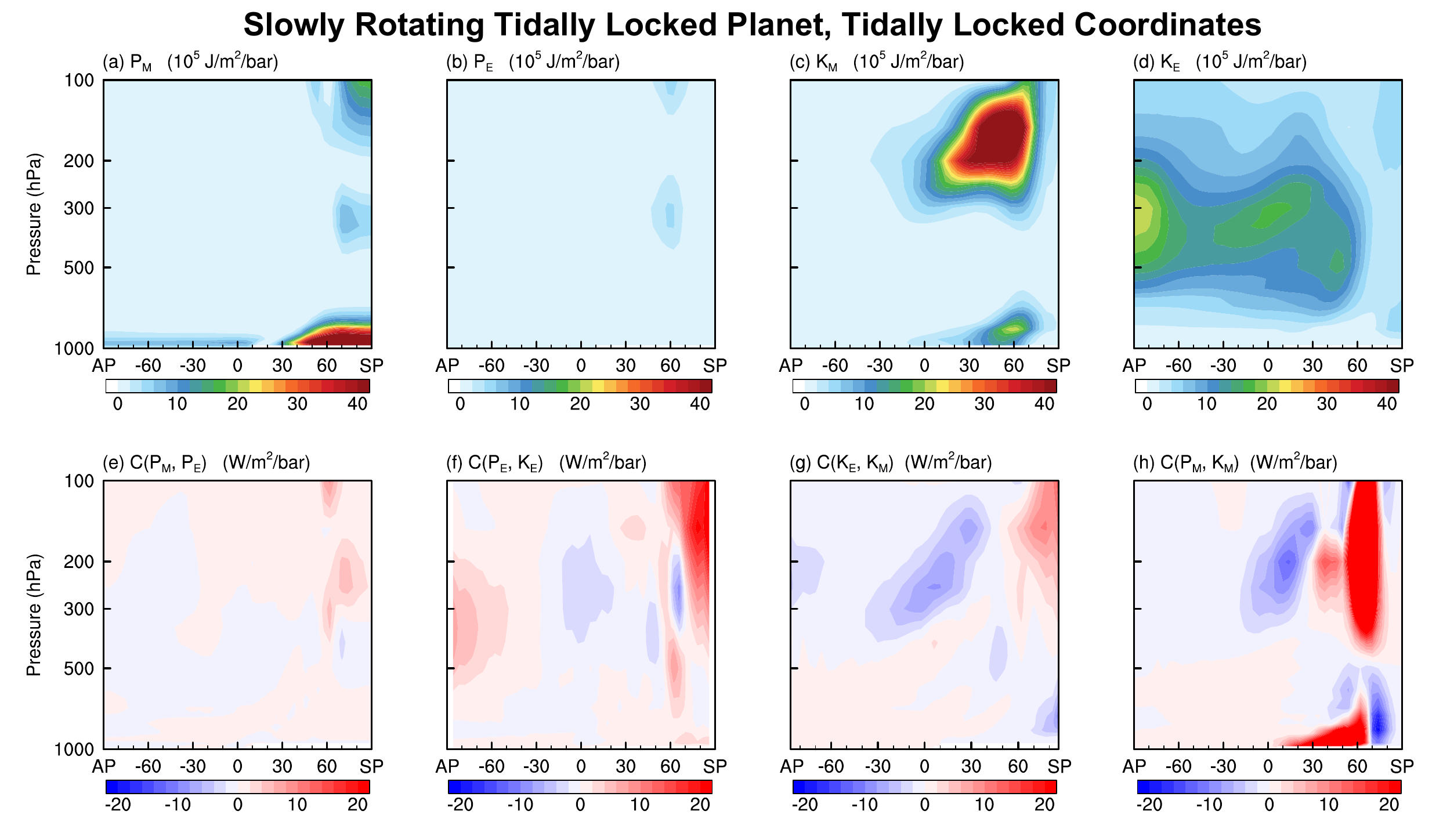}
    \caption{Same as Figure~\ref{fig:LEC_Earth} but on the slowly rotating tidally locked terrestrial planet. Panels are shown in the tidally locked coordinates. SP and AP are the substellar point and the antistellar point, respectively. The maximum in panel (a) is about $70\times10^5$ J\,m$^{-2}$\,bar$^{-1}$; the maximum in panel (c) is about $70\times10^5$ J\,m$^{-2}$\,bar$^{-1}$; the maximum in panel (h) is about $150$ W\,m$^{-2}$\,bar$^{-1}$.}
    \label{fig:LEC_Slow_TL_TL}
\end{figure*}

As a result of the WTGs in the free atmosphere (Figures~\ref{fig:climate_three_planets}(c) and \ref{fig:temp500_three_planets}(c)), P$_{\rm M}$ is nearly zero except very close to the surface around the substellar point (Figure~\ref{fig:LEC_Slow_TL_TL}(a)). Likewise, P$_{\rm E}$ is nearly zero throughout the atmosphere (Figure~\ref{fig:LEC_Slow_TL_TL}(b)). The horizontal flow in the upper branch of the global overturning circulation, known as zonal-mean meridional winds in the tidally locked coordinates (Figure~\ref{fig:climate_three_planets}(f)), contributes to the majority of K$_{\rm M}$ in the upper troposphere on the dayside (Figure~\ref{fig:LEC_Slow_TL_TL}(c)), while zonal-mean zonal winds in the tidally locked coordinates are very weak (Figure~\ref{fig:climate_three_planets}(c)). In addition, the backflow from the nightside to the substellar point contributes to the secondary maximum in K$_{\rm M}$ near the surface of the dayside, but with a relatively smaller wind speed due to surface friction. The stationary planetary waves contribute a portion of K$_{\rm E}$, which is centered on the nightside and extends to the dayside (Figure~\ref{fig:LEC_Slow_TL_TL}(d)).

Note that the superrotation is transformed to an eddy component in the tidally locked coordinates (Figure~\ref{fig:transform_zonal_jet}). Thus, the kinetic energy of the superrotation contributes a portion of K$_{\rm E}$ rather than K$_{\rm M}$. In general, K$_{\rm E}$ here is a measure of any wind that deviates from the global overturning circulation.

The conversion from P$_{\rm M}$ to P$_{\rm E}$ is inefficient throughout the atmosphere due to the WTGs and weak baroclinic activity (Figure~\ref{fig:LEC_Slow_TL_TL}(e)). P$_{\rm E}$ is converted to K$_{\rm E}$ on the nightside and around the substellar point, but K$_{\rm E}$ is converted back to P$_{\rm E}$ on the dayside, which is mainly related to the cross-isobaric motion of the equatorial superrotation (Figure~\ref{fig:LEC_Slow_TL_TL}(f)). The combine action of the conversion between P$_{\rm E}$ and K$_{\rm E}$ makes the global-mean conversion from P$_{\rm E}$ to K$_{\rm E}$ be inefficient. However, the ultimate reason is that the atmosphere is nearly barotropic. The conversion between K$_{\rm E}$ and K$_{\rm M}$ is dominant by K$_{\rm M}$ converting to K$_{\rm E}$, which occurs near the terminator (Figure~\ref{fig:LEC_Slow_TL_TL}(g)). It is caused by the interactions between the global overturning circulation and the eddy components, and the details are discussed in section~\ref{sec:compare_two_coords}. P$_{\rm M}$ is converted to K$_{\rm M}$ near the surface and in the upper troposphere around the substellar point, by the motion along the pressure gradient in the global overturning circulation (Figures~\ref{fig:climate_three_planets}(f) and \ref{fig:LEC_Slow_TL_TL}(h)). K$_{\rm M}$ is converted back to P$_{\rm M}$ in a limited region, and the net conversion is from P$_{\rm M}$ to K$_{\rm M}$. 

The structure of the conversion from P$_{\rm E}$ to K$_{\rm E}$ is common on a slowly rotating tidally locked planets. This is due to the fact that the crests of stationary Rossby waves usually lie near the eastern terminator and the troughs usually lie near the western terminator \citep[e.g.,][]{Carone_2015,Hammond_2018,Wang_2020}. That is, the equatorial superrotation acts against the pressure gradient force as it crosses the dayside, converting K$_{\rm E}$ to P$_{\rm E}$, and vice versa.

Comparing to Earth, the big difference is that P$_{\rm M}$ on the slowly rotating tidally locked planet is much smaller, which is mainly due to the WTGs caused by the smaller planetary rotation rate, i.e., 1/60 of that on Earth. Moreover, the strong and wide temperature inversion away from the substellar region makes the atmosphere be more stable than Earth and more difficult to generate motions, which also results in small P$_{\rm M}$. Another difference from Earth is that the energy conversion involved in baroclinic activity is inefficient, while the conversion from P$_{\rm M}$ to K$_{\rm M}$ is efficient, as a consequence of that a small planetary rotation rate makes the atmosphere be more barotropic and makes the thermal forcing tend to generate a strong circulation rather than temperature gradients \textcolor{black}{\citep[e.g.,][]{Edson_2011,Noda_2017,Komacek_2019}}.

\subsection{Comparison of LEC between standard and tidally locked coordinates}\label{sec:compare_two_coords}
\textcolor{black}{In this section, we compare the LEC between standard and tidally locked coordinates applied to rapidly and slowly rotating tidally locked planets. Briefly, the LEC in the standard coordinates is beneficial to describe P$_{\rm M}$ related to equator--pole temperature contrasts, K$_{\rm M}$ related to zonal-mean zonal winds (e.g., the equatorial superrotation), the conversion between P$_{\rm M}$ and K$_{\rm M}$, and wave--mean flow interactions. While the LEC in the tidally locked coordinates is beneficial to describe P$_{\rm M}$ related to day--night temperature contrasts, K$_{\rm M}$ related to the global overturning circulation, and the conversion between them. To gain more insight, we recalculate the LEC on the rapidly rotating tidally locked planet in the tidally locked coordinates and the LEC on the slowly rotating tidally locked planet in the standard coordinates, and compare them with the results in Sections~\ref{rapid} and \ref{slow}.}

\subsubsection{Comparison on rapidly rotating, tidally locked terrestrial planet}
\begin{figure*}
    \centering
    \includegraphics[width=0.9\textwidth]{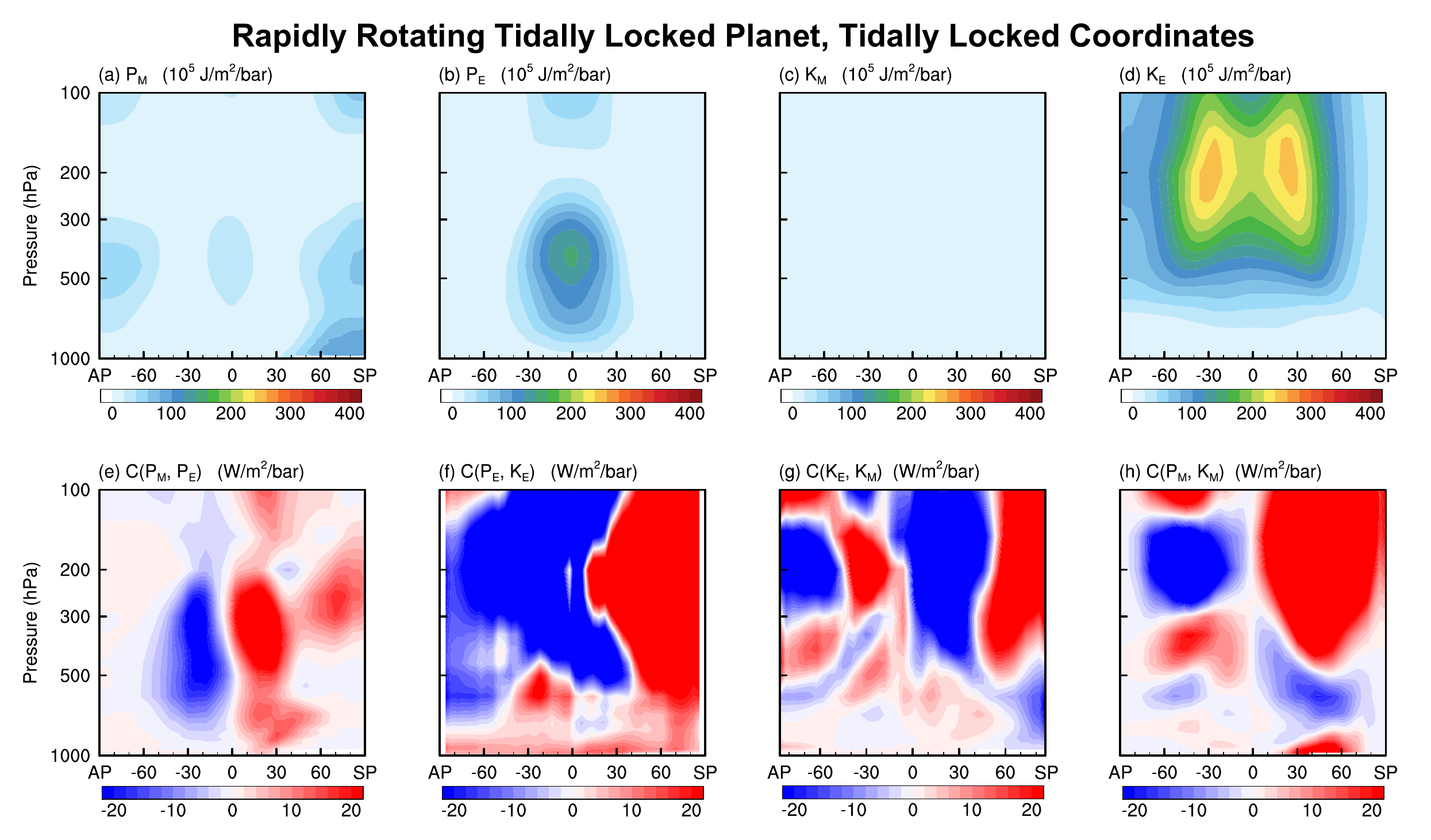}
    \caption{Same as Figure~\ref{fig:LEC_Earth} but on the rapidly rotating tidally locked terrestrial planet. Panels are shown in the tidally locked coordinates.}
    \label{fig:LEC_Fast_TL_TL}
\end{figure*}
\begin{figure*}
    \centering
    \includegraphics[width=0.9\textwidth]{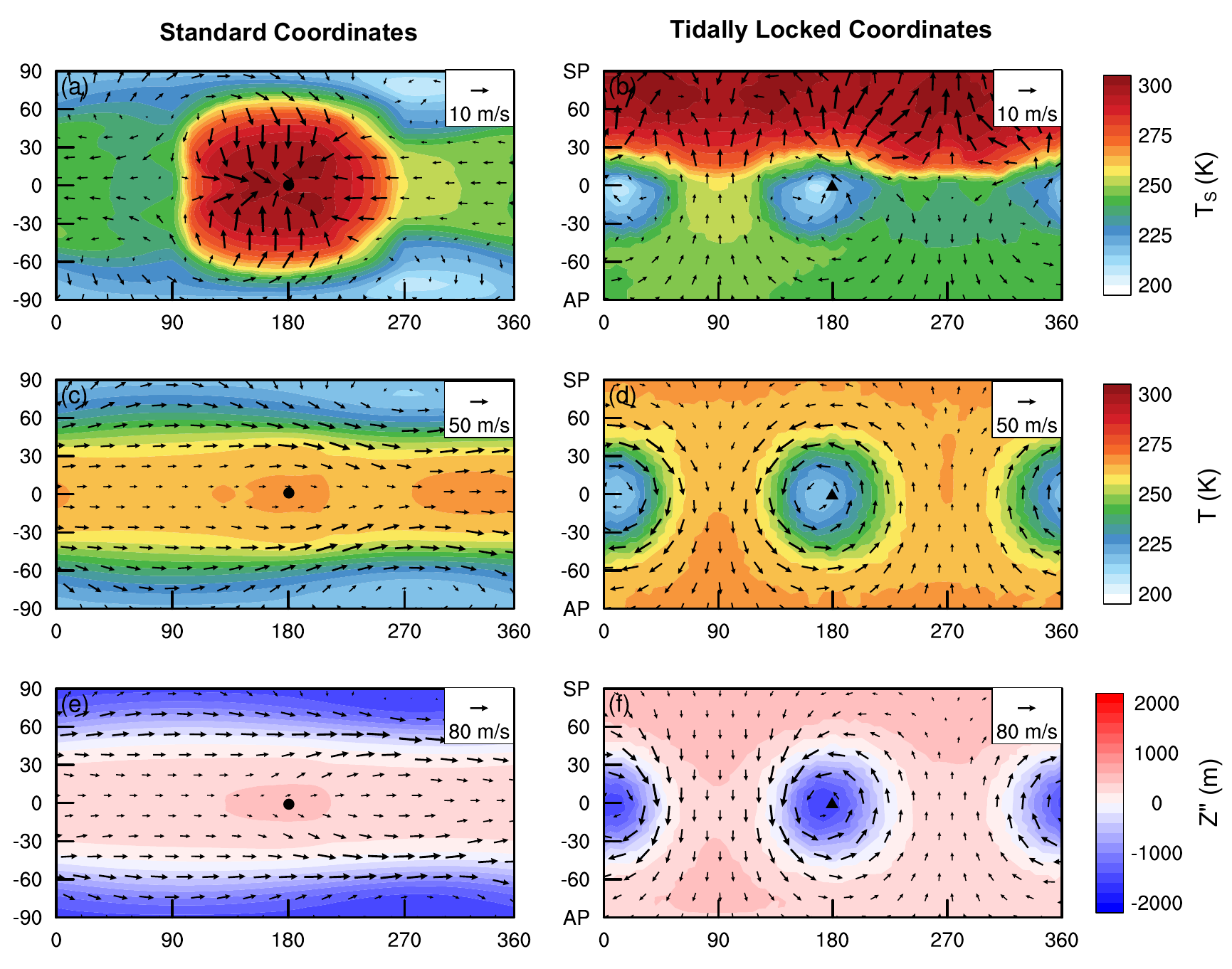}
    \caption{Climatic elements on the rapidly rotating tidally locked planet in the standard coordinates (left column) and in the tidally locked coordinates (right column). Upper row: surface temperature (contours) and the near-surface winds (vectors). Middle row: air temperature (contour) and winds (vector) at 500 hPa. Lower row: deviations of geopotential height from the global means (contour) and winds (vector) at 200 hPa. Black dot represents the substellar point, and black triangle represents the North Pole in the standard coordinates.}
    \label{fig:fast_climate}
\end{figure*}

\textcolor{black}{The global-mean vertical integrals of the LEC on the rapidly rotating tidally locked planet in the tidally locked coordinates are shown in Figure~\ref{fig:LEC_diagram_3}(b), and their structures are shown in Figure~\ref{fig:LEC_Fast_TL_TL}.}

\textcolor{black}{On a rapidly rotating tidally locked planet, the air temperature and geopotential height are zonally homogeneous in the free atmosphere, and the sharp contrasts are between the equatorial and polar regions (Figures~\ref{fig:fast_climate}(c) and (e)). The winds in the free atmosphere also behave as the zonal jets. These equator--pole contrasts become eddy components in the tidally locked coordinates, and so do the zonal jets (Figures~\ref{fig:fast_climate}(d) and (f)). Thus, P$_{\rm E}$ and K$_{\rm E}$ in the tidally locked coordinates are much larger than those in the standard coordinates, respectively, which may lead a misconception that eddies dominate this planet rather than large-scale circulation (Figure~\ref{fig:LEC_diagram_3}). Temperature and winds exhibit a day--night asymmetry only very close to the surface (Figure~\ref{fig:fast_climate}(a)), so that P$_{\rm M}$ and K$_{\rm M}$ in the tidally locked coordinates are smaller than those in the standard coordinates and centered only close to the surface (Figures~\ref{fig:LEC_Fast_TL_TL}(a) and (c)).}

\textcolor{black}{Since both zonal-mean zonal winds and waves belong to eddy components in the tidally locked coordinates, the conversion between K$_{\rm E}$ and K$_{\rm M}$ here no longer describes the wave--mean flow interactions. The structure of this conversion rate is complicated and atypical, and the corresponding dynamical process is unclear on this planet (Figure~\ref{fig:LEC_Fast_TL_TL}(g)). The zonal-mean meridional wind speed in the tidally locked coordinates is generally larger than that in the standard coordinates, because meridional winds in the standard coordinates are reversed between the day and night hemispheres. For example, the former is over 10 m\,s$^{-1}$ on this planet, while the latter is 4 m\,s$^{-1}$. Thus, the conversion from P$_{\rm M}$ to K$_{\rm M}$ in the tidally locked coordinates is more efficient than that in the standard coordinates (Figure~\ref{fig:LEC_diagram_3}).}

\textcolor{black}{The integrals and cross-sections of energy and conversion rates in the tidally locked coordinates can be understood from the distributions of air temperature, geopotential height, and horizontal winds. However, they are more complicated and lead to less insight into the atmospheric circulation on this planet than those in the standard coordinates, so we omit them here.}

\textcolor{black}{Note that the total potential energy, i.e., the sum of P$_{\rm M}$ and P$_{\rm E}$, is about $44.8\times10^5$ J\,m$^{-2}$ and is the same in both two coordinates (Figure~\ref{fig:LEC_diagram_3}). Likewise, the total kinetic energy is about $78.4\times10^5$ J\,m$^{-2}$ and is also the same in the two coordinates. They do not depend on the choice of the coordinates.}

\subsubsection{Comparison on slowly rotating, tidally locked terrestrial planet}
\begin{figure*}
    \centering
    \includegraphics[width=0.9\textwidth]{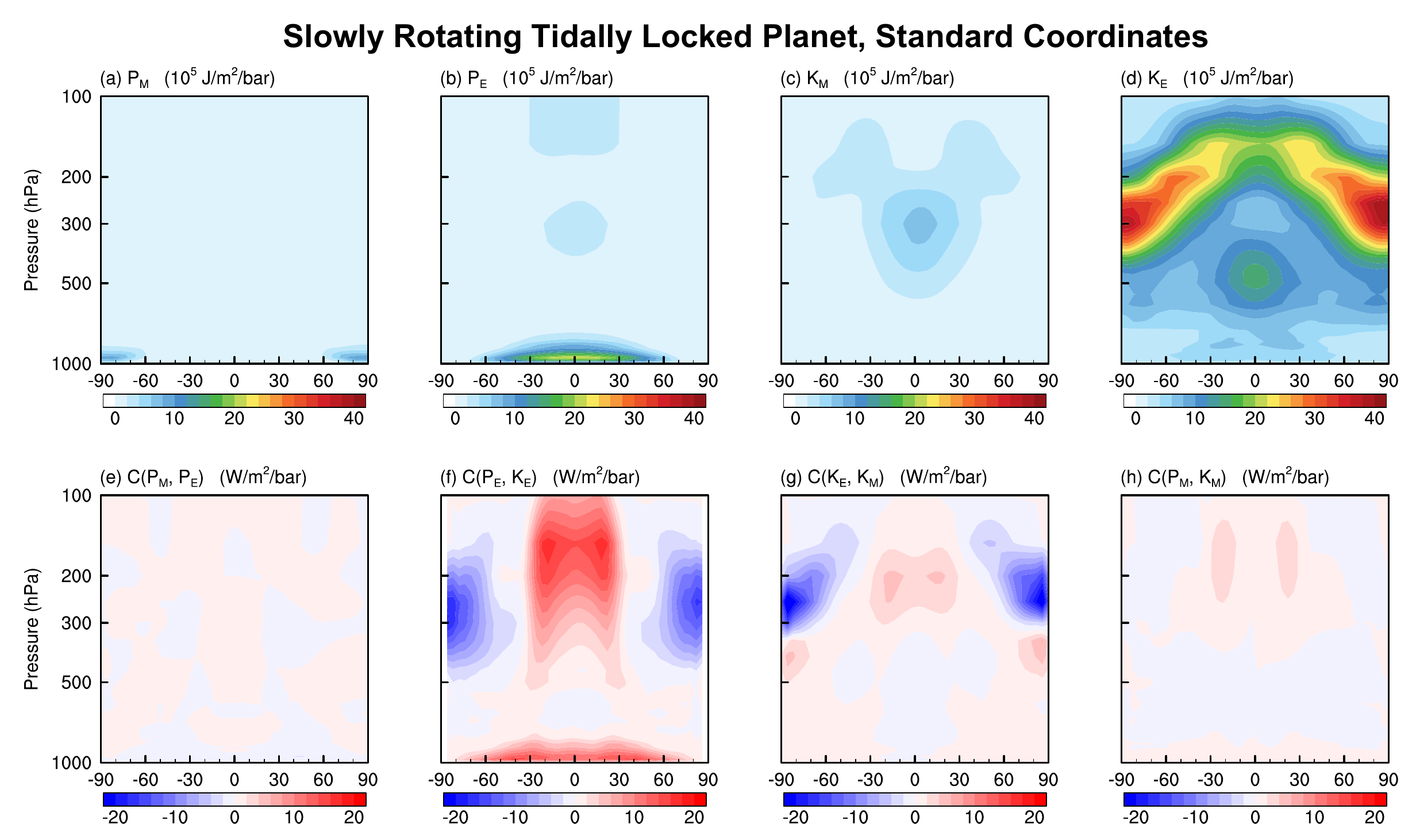}
    \caption{Same as Figure~\ref{fig:LEC_Earth} but on the slowly rotating tidally locked terrestrial planet. Panels are shown in the standard coordinates.}
    \label{fig:LEC_Slow_TL_ST}
\end{figure*}
\begin{figure*}
    \centering
    \includegraphics[width=0.9\textwidth]{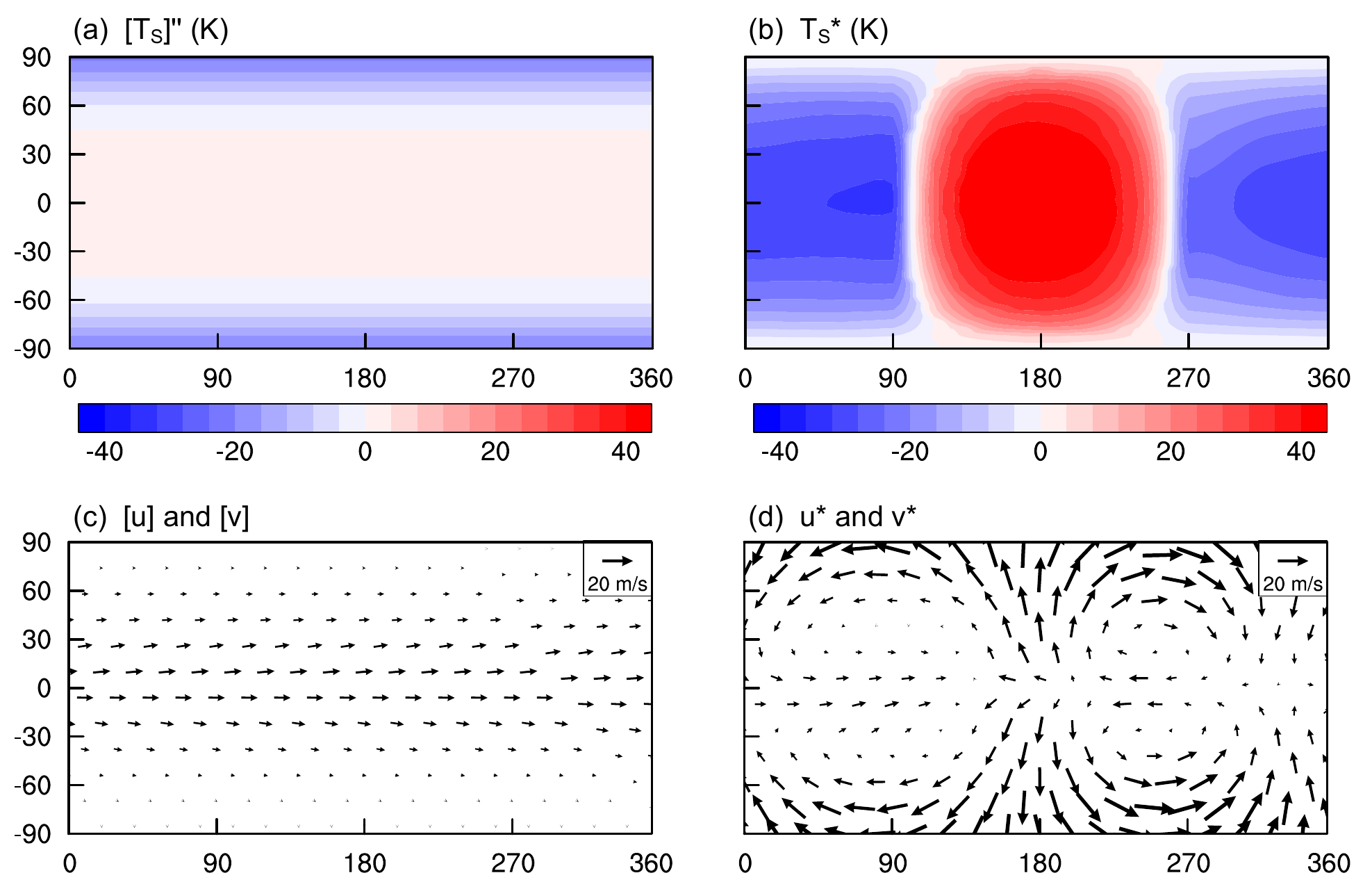}
    \caption{Decomposition of temperatures and winds on the slowly rotating tidally locked terrestrial planet in the standard coordinates. (a) Meridional contrast of zonal-mean surface temperature; (b) eddy surface temperature; (c) zonal-mean zonal and meridional winds at 300 hPa; (d) eddy winds at 300 hPa.}
    \label{fig:mean_and_eddy}
\end{figure*}

\textcolor{black}{The global-mean vertical integrals of the LEC on the slowly rotating tidally locked planet in the standard coordinates are shown in Figure~\ref{fig:LEC_diagram_2}(b), and their structures are shown in Figure~\ref{fig:LEC_Slow_TL_ST}.}

\textcolor{black}{On a slowly rotating tidally locked planet,} the day--night temperature contrast is usually larger than the meridional contrast. For example, the day--night surface temperature contrast in our experiment is larger than 80 K, while the meridional contrast is smaller than 30 K (Figures~\ref{fig:mean_and_eddy}(a) and (b)). Likewise, the wind speeds of the global overturning circulation are larger than those of the zonal-mean zonal winds, for example, about 20 m\,s$^{-1}$ for the former and about 5 m\,s$^{-1}$ for the latter at 300 hPa in our experiment (Figures~\ref{fig:mean_and_eddy}(c) and (d)). Thus, in contrast to the LEC in the tidally locked coordinates, P$_{\rm M}$ and K$_{\rm M}$ are smaller in the standard coordinates, but P$_{\rm E}$ and K$_{\rm E}$ are larger than those in the tidally locked coordinates (Figure~\ref{fig:LEC_diagram_2}), \textcolor{black}{as the day--night temperature contrast and the overturning circulation belong to eddy components here (Figures~\ref{fig:LEC_Slow_TL_ST}(b) and (d)).}

\textcolor{black}{The main conversion path in the standard coordinates} becomes P$_{\rm E}$ to K$_{\rm E}$, which may lead to a misconception that the conversion is dominated by the baroclinic activity (Figure~\ref{fig:LEC_Slow_TL_ST}(f)). Indeed, the main conversion occurs in the large-scale dynamical process, where the day--night temperature contrast induced by the uneven stellar radiation generates large-scale upwellings and downwellings. \textcolor{black}{However, the meaning of the conversion between K$_{\rm E}$ and K$_{\rm M}$ in the standard coordinates is clear. Its structure is consistent with wave--mean flow interactions, where momentum is transported to the equator by waves to maintain the equatorial superrotation (Figure~\ref{fig:LEC_Slow_TL_ST}(g)).} \textcolor{black}{By contrast, the conversion between K$_{\rm E}$ and K$_{\rm M}$ in the tidally locked coordinates is complex. On this planet,} the tidally locked zonal-mean zonal winds are nearly zero, but the tidally locked zonal-mean meridional winds are strong (Figures~\ref{fig:climate_three_planets}(c) and (f)), so that the conversion between K$_{\rm E}$ and K$_{\rm M}$ in the tidally locked coordinates is more related to the shear instability of the meridional winds, i.e., the global overturning circulation (see Equation~(\ref{eq:C(KE,KM)}) in Appendix~\ref{appendix_b}). In our calculations, the bulk of the conversion from K$_{\rm M}$ to K$_{\rm E}$ is contributed by the horizontal shear of the global overturning circulation, which causes the accumulation of the large-scale momentum and the conversion to the vortex momentum  (see the second term in the right-hand side in Equation~(\ref{eq:C(KE,KM)}), figure not shown).

\textcolor{black}{Note that the total potential energy is about $2.8\times10^5$ J\,m$^{-2}$, the total kinetic energy is about $9.9\times10^5$ J\,m$^{-2}$, and they are the same in the two coordinates, respectively (Figure~\ref{fig:LEC_diagram_2}).}

\section{Summary and Discussions}\label{sec:summary}
In this study, we employ the Lorenz energy cycle (LEC) to understand the atmospheric circulation on tidally locked terrestrial planets. We use ExoCAM to simulate atmospheric circulation on both rapidly and slowly rotating tidally locked terrestrial planets, calculate their LECs, and compare them with that on Earth. The main conclusions are as follows:     

\begin{enumerate}
    \renewcommand{\labelenumi}{(\theenumi)}
      \item On the rapidly rotating tidally locked planet, the mean potential energy P$_{\rm M}$ and eddy potential energy P$_{\rm E}$ are comparable to those on Earth, because both them have similar steep meridional temperature gradients. The mean kinetic energy K$_{\rm M}$ is much larger than that on Earth, mainly related to much larger wind speeds. The two paths of energy conversion, P$_{\rm M}\rightarrow$P$_{\rm E}\rightarrow$K$_{\rm E}$ and P$_{\rm M}\rightarrow$K$_{\rm M}\rightarrow$K$_{\rm E}$, are both effective. The former path is mainly associated with baroclinic instabilities, and the latter path is associated with large-scale thermal-driven circulation and barotropic instabilities. These suggest that the atmosphere of rapidly rotating tidally locked planets is in a mixed dynamical regime of single-cell circulation and baroclinic eddies.
      
      \item On the slowly rotating tidally locked planet, P$_{\rm M}$ and P$_{\rm E}$ are small. This is because the slow rotation rate makes the planet be in a weak temperature gradient regime. Meanwhile, \textcolor{black}{the temperature inversion that lies over the entire nightside and part of the dayside makes the atmosphere be very stable}, also contributes to the small potential energy. K$_{\rm M}$ and K$_{\rm E}$ are comparable to those on Earth. However, in the tidally locked coordinates, K$_{\rm M}$ is the measure of the global overturning circulation, and K$_{\rm E}$ is the measure of waves and zonal-mean zonal jets including the equatorial superrotation. The main path of energy conversion is P$_{\rm M}\rightarrow$K$_{\rm M}\rightarrow$K$_{\rm E}$, associate with cross-isobaric motions in the global overturning circulation and interactions between the global overturning circulation and eddy components.
   \end{enumerate}

\textcolor{black}{In this study, the main factor discussed to affect the LEC is the planetary rotation rate. Although only two experiments have been performed, the rough tendency is that air temperature gradients and wind speed decrease as the rotation rate becomes smaller, as can be found in previous studies \citep[e.g.,][]{Carone_2015,Carone_2016,Noda_2017}. As a result, the total available energy stored in the atmosphere is less on planets with slower rotation. Meanwhile, a smaller planetary rotation rate makes the atmosphere less baroclinic \citep{Komacek_2019}. Thus, the conversion paths associated with baroclinic instabilities are less efficient, and the conversion between P$_{\rm M}$ and K$_{\rm M}$ becomes dominant.}

\textcolor{black}{The planetary rotation rate also determines the ratio of the Rossby deformation radius to the planetary radius, and hence a regime transition in the atmosphere \citep[e.g.,][]{Carone_2015,Haqq_Misra_2018}. On a slow rotator, the atmospheric state has a large day--night asymmetry, and a LEC in the tidally locked coordinates is more appropriate.
Since the planet becomes a rapid rotator, the atmospheric state tends to be zonally homogeneous, so that a LEC in the standard coordinates is more appropriate. For an Earth-like tidally locked planet, the rotation period that separates the rapid and slow rotators is usually 5--10 Earth days \citep{Edson_2011,Yang_2014,Noda_2017}. However, both forms of LEC have shortcomings, as the LEC in the standard coordinates may not be robust very close to the surface, and the LEC in the tidally locked coordinates is disabled in describing wave--mean flow interactions. There may be a new form of LEC that can be applied to both rapid and slow rotators, which needs to be investigated in the future.}

\textcolor{black}{Other factors may also affect the LEC, but are not included in this study.} For example, both day--night temperature contrast and wind speeds monotonically decrease as the background air pressure is increased \citep{Kite_2011,Leconte_2013,Wordsworth_2015,Zhang_2020}. They would make the available potential and kinetic energy per unit mass of air decrease, but the change of the total atmospheric energy is unclear, as the total mass of the atmosphere is increased. Atmospheric compositions should also have impacts on the LEC. \cite{Ding_2019} showed that excluding greenhouse gases from the atmosphere would decrease the day--night temperature contrast and wind speeds; \cite{Wang_2022} showed that the atmospheric circulation in pure N$_2$ atmosphere without any greenhouse gas is very weak. They suggest that the potential and kinetic energy would be smaller in an atmosphere without any greenhouse gas.

In this study, we do not calculate the generation rates of potential energy and the dissipation rates of kinetic energy, which are inevitable to close the LEC. However, they could be estimated from the remnants of four conversion rates by assuming an equilibrium state \citep[e.g.,][]{Peixoto_1992,Li_2007,Pan_2017}. By doing so, our calculations obtain the total generation rates of potential energy (= total dissipation rates of kinetic energy) with values of 2.61, 4.24, and 2.39 W\,m$^{-2}$ on the three planets, respectively. Moreover, the generation and dissipation rates can be used to obtain a more realistic efficiency of the atmospheric heat engine, \textcolor{black}{which is regarded as the ratio of the total dissipation rate of kinetic energy to the mean net incoming solar radiation}. \textcolor{black}{On Earth, the total dissipation rate is 2.61 W\,m$^{-2}$ and the mean net incoming solar radiation is 238 W\,m$^{-2}$. They lead to $\eta\approx1.1\%$, which is much smaller than the ideal limit based on the Carnot's heat engine, $\eta\approx10\%$ \citep{Peixoto_1992}.} Likewise, the LEC will also provide a smaller efficiency than the ideal limit on tidally locked planets, for example, $2.3\%$ \textcolor{black}{(4.24 W\,m$^{-2}$/187 W\,m$^{-2}$)} for the rapidly rotating planet and $1.4\%$ \textcolor{black}{(2.39 W\,m$^{-2}$/175 W\,m$^{-2}$)} for the slowly rotating planet in our experiments. \cite{Koll_2016} estimated the surface wind speed on tidally locked planets by using the ideal efficiency of atmospheric heat engine, and the efficiency we obtained may help to optimize their estimations.          

The formulas of the LEC in this study is only applied to atmospheres on terrestrial planets. Based on the shallow atmosphere approximation, we do not consider the kinetic energy of vertical motion or the conversion rates contributed from corresponding metric terms (e.g., \textcolor{black}{$-u_{TL}\omega_{TL}/a$} in Equation~(\ref{dudt_TL})). However, this approximation may be invalid when the vertical motion is comparable to the horizontal motion. That means that the primitive equations and the LEC in this study may be inapplicable for gas planets. They must be corrected by considering all the terms involved in the vertical motion. \textcolor{black}{The application of the LEC to gas giants is useful to estimate the efficiency of the planetary heat engine, which is different from that of a terrestrial planet. In solar system, gas giants (e.g., Jupiter and Saturn) receive less solar radiation than terrestrial planets (e.g., Venus and Earth) but hold stronger winds, suggesting a more efficient planetary heat engine, which may be due to the absence of solid surface or atmospheric compositions \citep{Showman_2009,Ingersoll_2013}. This estimate may be beneficial for predicting wind speeds including superrotation and for quantifying the energy sources of jet streams on gas giants. Moreover, the LEC may be a useful way to evaluate various models for gas giants by comparing the results computed from these models. }     

\begin{appendix}
\section{transformation relation}\label{appendix_a}
In order to transform the physical quantities in the standard coordinates to their counterparts in the tidally locked coordinates, we derive a transformation relation between the two coordinates. This relation has been derived by \cite{Koll_2015}, but with a left-hand coordinate system (i.e., $\mathbf{\hat{e}_i}\times\mathbf{\hat{e}_j}=-\mathbf{\hat{e}_k}$, see their Figure 1(b)). Here we make sure that the two coordinates are right-hand systems (i.e., $\mathbf{\hat{e}_i}\times\mathbf{\hat{e}_j}=\mathbf{\hat{e}_k}$). We put the substellar point at latitude/longitude $\left(\phi,\lambda\right)=\left(0^{\circ},180^{\circ}\right)$ in the standard coordinates and at tidally locked latitude $\phi_{TL}=90^{\circ}$ in the tidally locked coordinates, and also put the South and North Poles in the standard coordinates at $\left(\phi_{TL},\lambda_{TL}\right)=\left(0^{\circ},0^{\circ}\right)$ and $\left(\phi_{TL},\lambda_{TL}\right)=\left(0^{\circ},180^{\circ}\right)$, respectively (Figure~\ref{fig:changing_coordinates}). We transform the two coordinates into the Cartesian coordinates, where x-axis links the spherical core and the substellar point, and z-axis links the spherical core and the North Pole in the standard coordinates, so that
\begin{eqnarray}\label{eq:standard_to_xyz}
    x&=&-r\cos\lambda\cos\phi,\nonumber\\
    y&=&-r\sin\lambda\cos\phi,\nonumber\\
    z&=&r\sin\phi,
\end{eqnarray}
and
\begin{eqnarray}\label{eq:tidally_to_xyz}
    x&=&r\sin\phi_{TL},\nonumber\\
    y&=&r\sin\lambda_{TL}\cos\phi_{TL},\nonumber\\
    z&=&-r\cos\lambda_{TL}\cos\phi_{TL},
\end{eqnarray}
where $r$ is the radial distance from the center of the planet. Combining Equations~(\ref{eq:standard_to_xyz}) and (\ref{eq:tidally_to_xyz}) yields the transformation relations between the two coordinates:
\begin{eqnarray}\label{eq:coordinate_relations}
    \lambda_{TL}&=&\tan^{-1}\left(\frac{\sin\lambda}{\tan\phi}\right),\nonumber\\
    \phi_{TL}&=&\sin^{-1}\left(-\cos\lambda\cos\phi\right),\nonumber\\
    \lambda&=&\tan^{-1}\left(\frac{\sin\lambda_{TL}}{\tan\phi_{TL}}\right), \nonumber\\
    \phi&=&\sin^{-1}\left(-\cos\lambda_{TL}\cos\phi_{TL}\right),
\end{eqnarray}
where $\lambda$ and $\lambda_{TL}$ belong to $[0,2\pi]$, and $\phi$ and $\phi_{TL}$ belong to $[-\pi/2,\pi/2]$. Because $\cos\phi\geq0$ and $\cos\phi_{TL}\geq0$, Equations~(\ref{eq:standard_to_xyz}) and (\ref{eq:tidally_to_xyz}) also yield the sign relations:
\begin{eqnarray}\label{eq:sign_relations}
    \frac{\cos\lambda}{\left|\cos\lambda\right|}&=&-\frac{\sin\phi_{TL}}{\left|\sin\phi_{TL}\right|}=-\frac{\tan\phi_{TL}}{\left|\tan\phi_{TL}\right|},\nonumber\\
    \frac{\sin\lambda}{\left|\sin\lambda\right|}&=&-\frac{\sin\lambda_{TL}}{\left|\sin\lambda_{TL}\right|},
\end{eqnarray}
which are used to calculate the transformation of trigonometric functions.

The transformation relations for the scalars including the temperature and the geopotential height follow the transformation of the coordinates. That is, the value of a scalar in the tidally locked coordinates is equal to the value at the corresponding longitude and latitude in the standard coordinates, i.e.,
\begin{equation}       
A(\lambda_{TL},\phi_{TL})=A(\lambda(\lambda_{TL},\phi_{TL}),\phi(\lambda_{TL},\phi_{TL})),
\end{equation}
where the transformation of the coordinates is from Equation~(\ref{eq:coordinate_relations}).

The winds in the standard coordinates are $(u,v,w)\equiv(r\cos\phi(D\lambda/Dt),r(D\phi/Dt),Dr/Dt)$. Likewise, the winds in the tidally locked coordinates are
\begin{eqnarray}\label{eq:wind_relations}
    u_{TL}&\equiv&r\cos\phi_{TL}\frac{D\lambda_{TL}}{Dt}\nonumber\\
    &=&r\cos\phi_{TL}\left(\frac{\partial\lambda_{TL}}{\partial\lambda}\frac{D\lambda}{Dt}+\frac{\partial\lambda_{TL}}{\partial\phi}\frac{D\phi}{Dt}\right)\nonumber\\
    &=&\cos\phi_{TL}\left(\frac{\partial\lambda_{TL}}{\partial\lambda}\frac{u}{\cos\phi}+\frac{\partial\lambda_{TL}}{\partial\phi}v\right),\nonumber\\
    v_{TL}&\equiv&r\frac{D\phi_{TL}}{Dt}\nonumber\\
    &=&r\left(\frac{\partial\phi_{TL}}{\partial\lambda}\frac{D\lambda}{Dt}+\frac{\partial\phi_{TL}}{\partial\phi}\frac{D\phi}{Dt}\right)\nonumber\\
    &=&\frac{\partial\phi_{TL}}{\partial\lambda}\frac{u}{\cos\phi}+\frac{\partial\phi_{TL}}{\partial\phi}v,\nonumber\\
    w_{TL}&=&\frac{Dr}{Dt}=w.
\end{eqnarray}
Substituting Equation~(\ref{eq:coordinate_relations}) into Equation~(\ref{eq:wind_relations}), we obtain
\begin{equation}\label{eq:wind_relations_matrix}
    \left[\begin{array}{lll}u_{TL} \\v_{TL} \\w_{TL}\end{array}\right]=\left[\begin{array}{lll}\frac{\cos\lambda\sin\phi}{\sqrt{1-\cos^2\lambda\cos^2\phi}} & \frac{-\sin\lambda}{\sqrt{1-\cos^2\lambda\cos^2\phi}} & 0 \\ \frac{\sin\lambda}{\sqrt{1-\cos^2\lambda\cos^2\phi}} & \frac{\cos\lambda\sin\phi}{\sqrt{1-\cos^2\lambda\cos^2\phi}} & 0 \\ 0 & 0 & 1\end{array}\right]\left[\begin{array}{lll}u \\v \\w\end{array}\right].
\end{equation}
\textcolor{black}{The general transformation procedure is to calculate $u_{TL}$ and $v_{TL}$ at an arbitrary point in the standard coordinates using Equation~(\ref{eq:wind_relations_matrix}), and then to determine the position of the point in the tidally locked coordinates using Equation~(\ref{eq:coordinate_relations}). Substituting Equations~(\ref{eq:coordinate_relations}) and (\ref{eq:sign_relations}) into the transformation matrix of the winds yields
\begin{eqnarray}
    \frac{\cos\lambda\sin\phi}{\sqrt{1-\cos^2\lambda\cos^2\phi}}&=&-\frac{\left|\cos\lambda\right|\sin\phi}{\sqrt{1-\cos^2\lambda\cos^2\phi}}\frac{\tan\phi_{TL}}{\left|\tan\phi_{TL}\right|} \nonumber\\
    &=&-\frac{\sin\phi}{\sqrt{\tan^2\lambda+\sin^2\phi}}\frac{\tan\phi_{TL}}{\left|\tan\phi_{TL}\right|} \nonumber\\
    &=&\frac{\cos\lambda_{TL}\cos\phi_{TL}\tan\phi_{TL}}{\sqrt{\sin^2\lambda_{TL}+\cos^2\lambda_{TL}\sin^2\phi_{TL}}} \nonumber\\
    &=&\frac{\cos\lambda_{TL}\tan\phi_{TL}}{\sqrt{\sin^2\lambda_{TL}+\tan^2\phi_{TL}}}, \nonumber
\end{eqnarray}
\begin{eqnarray}
    \frac{\sin\lambda}{\sqrt{1-\cos^2\lambda\cos^2\phi}}&=&-\frac{\sin\lambda_{TL}}{\left|\sin\lambda_{TL}\right|}\frac{\left|\sin\lambda\right|}{\sqrt{1-\cos^2\lambda\cos^2\phi}}\nonumber\\
    &=&-\frac{\sin\lambda_{TL}}{\left|\sin\lambda_{TL}\right|}\frac{1}{\sqrt{1+\cot^2\lambda\sin^2\phi}} \nonumber\\
    &=&-\frac{\sin\lambda_{TL}}{\sqrt{\sin^2\lambda_{TL}+\cos^2\lambda_{TL}\sin^2\phi_{TL}}} \nonumber\\
    &=&-\frac{\sin\lambda_{TL}\sec\phi_{TL}}{\sqrt{\sin^2\lambda_{TL}+\tan^2\phi_{TL}}}, \nonumber
\end{eqnarray}
so that Equation~(\ref{eq:wind_relations_matrix}) can also be expressed in terms of the tidally locked coordinates,
\begin{equation}\label{eq:wind_relations_matrix_tl}
    \left[\begin{array}{lll}u_{TL} \\v_{TL} \\w_{TL}\end{array}\right]=\left[\begin{array}{lll}\frac{\cos\lambda_{TL}\tan\phi_{TL}}{\sqrt{\sin^2\lambda_{TL}+\tan^2\phi_{TL}}} & \frac{\sin\lambda_{TL}\sec\phi_{TL}}{\sqrt{\sin^2\lambda_{TL}+\tan^2\phi_{TL}}} & 0 \\ -\frac{\sin\lambda_{TL}\sec\phi_{TL}}{\sqrt{\sin^2\lambda_{TL}+\tan^2\phi_{TL}}} & \frac{\cos\lambda_{TL}\tan\phi_{TL}}{\sqrt{\sin^2\lambda_{TL}+\tan^2\phi_{TL}}} & 0 \\ 0 & 0 & 1\end{array}\right]\left[\begin{array}{lll}u \\v \\w\end{array}\right],
\end{equation}
which can be used to directly calculate the transformation of some special wind fields in the standard coordinates, such as a uniform zonal-mean zonal winds.} It is easy to prove that
\begin{eqnarray}
    u_{TL}^2+v_{TL}^2+w_{TL}^2=u^2+v^2+w^2,\nonumber
\end{eqnarray}
so that the transformation between the standard and the tidally locked coordinates satisfies the conservation of energy.

\section{formulas of LEC}\label{appendix_b}
The formulas of the LEC in the standard coordinates have been derived by \cite{Peixoto_1974,Peixoto_1992}. We follow their method and obtain the LEC in the tidally locked coordinates. Since the primitive equations in the standard and the tidally locked coordinates are the same, the formulas of the LEC in the two coordinates are also the same, albeit with different details. For example, $u$ is the winds along the longitude in the standard coordinates but along the tidally locked longitude in the tidally locked coordinates. The zonal mean is the average along the longitude in the standard coordinates, but is the average along the tidally locked longitude in the tidally locked coordinates, so do the deviations from zonal means. The formulas and the descriptions of the LEC are shown below.
\begin{enumerate}
    \item[$\bullet$]Mean available potential energy
    \begin{equation}\label{eq:PM}
        {\rm P_M}=\frac{c_p}{2}\int\gamma\left[\bar{T}\right]^{''2}dm,
    \end{equation}
    which mainly depends on the departures of zonal-mean isotherms from their global means. The zonal-mean isotherms are always tilted by large-scale uneven heating or cooling, while their global means are horizontal (Figure~\ref{fig:LEC_framework}(b)).    
    \item[$\bullet$]Eddy available potential energy
    \begin{equation}\label{eq:PE}
        {\rm P_E}=\frac{c_p}{2}\int\gamma\left[\overline{T^{'2}}+\bar{T}^{*2}\right]dm,
    \end{equation}
    which mainly depends on the temporal variability of temperatures and the departures from temporal- and zonal-mean temperatures. P$_{\rm E}$ is also generated by uneven heating or cooling, but usually in smaller scales.
    \item[$\bullet$]Mean kinetic energy
    \begin{equation}\label{eq:KM}
        {\rm K_M}=\frac{1}{2}\int\left(\left[\bar{u}\right]^2+\left[\bar{v}\right]^2\right)dm,
    \end{equation}
    which depends on the strengths of jet streams and overturning circulation.
    \item[$\bullet$]Eddy kinetic energy
    \begin{equation}\label{eq:KE}
        {\rm K_E}=\frac{1}{2}\int\left[\overline{u^{'2}}+\overline{v^{'2}}+\bar{u}^{*2}+\bar{v}^{*2}\right]dm,
    \end{equation}
    which depends on the wind speeds of transient and stationary eddies.
    \item[$\bullet$]Conversion rate from P$_{\rm M}$ to P$_{\rm E}$
    \begin{eqnarray}\label{eq:C(PM,PE)}
        {\rm C\left(P_M,P_E\right)}=-\int c_p\gamma\left[\overline{v^{'}T^{'}}+\bar{v}^{*}\bar{T}^{*}\right]\frac{\partial\left[\bar{T}\right]}{a\partial\phi}dm-\int c_pp^{-\kappa}\left[\overline{\omega^{'}T^{'}}+\bar{\omega}^{*}\bar{T}^{*}\right]\frac{\partial}{\partial p}\left(\gamma p^{\kappa}\left[\bar{T}\right]^{''}\right)dm,
    \end{eqnarray}
    which mainly depends on the heat transport by baroclinic eddies. In this process, the isotherms in the longitude-latitude cross-section are warped, which reduces the zonal-mean meridional temperature gradients but leads to additional variance in the longitude direction, equivalent to converting P$_{\rm M}$ to P$_{\rm E}$ (Figure~\ref{fig:LEC_framework}(c)). 
    \item[$\bullet$]Conversion rate from P$_{\rm E}$ to K$_{\rm E}$
    \begin{eqnarray}\label{eq:C(PE,KE)}
        {\rm C\left(P_E,K_E\right)}=-\int g\left[\overline{\frac{u^{'}\partial Z^{'}}{a\cos\phi\partial\lambda}}+\overline{\frac{v^{'}\partial Z^{'}}{a\partial\phi}}+\frac{\bar{u}^{*}\partial\bar{Z}^{*}}{a\cos\phi\partial\lambda}+\frac{\bar{v}^{*}\partial\bar{Z}^{*}}{a\partial\phi}\right]dm,
    \end{eqnarray}
    which depends on the cross-isobaric motions in eddies. The pressure gradient force would do work on the air parcel which moves along the pressure gradients, and convert P$_{\rm E}$ to K$_{\rm E}$. Otherwise, the air parcel moving against the pressure gradient force would convert K$_{\rm E}$ to P$_{\rm E}$ (Figure~\ref{fig:LEC_framework}(d)).
    \item[$\bullet$]Conversion rate from K$_{\rm E}$ to K$_{\rm M}$
    \begin{eqnarray}\label{eq:C(KE,KM)}
        {\rm C\left(K_E,K_M\right)}&=&\int\left[\overline{v^{'}u^{'}}+\bar{v}^{*}\bar{u}^{*}\right]\cos\phi\frac{\partial\left[\bar{u}\right]/\cos\phi}{a\partial\phi}dm+\int\left[\overline{v^{'2}}+\bar{v}^{*2}\right]\frac{\partial\left[\bar{v}\right]}{a\partial\phi}dm \nonumber \\
        &&+\int\left[\overline{\omega^{'}u^{'}}+\bar{\omega}^{*}\bar{u}^{*}\right]\frac{\partial\left[\bar{u}\right]}{\partial p}dm+\int\left[\overline{\omega^{'}v^{'}}+\bar{\omega}^{*}\bar{v}^{*}\right]\frac{\partial\left[\bar{v}\right]}{\partial p}dm \nonumber \\
        &&-\int\left[\overline{u^{'2}}+\bar{u}^{*2}\right]\left[\bar{v}\right]\frac{\tan\phi}{a}dm,
    \end{eqnarray}
    which normally depends on the wave--mean flow interactions. Under a positive $\beta$-plane, eddies \textcolor{black}{with group velocity directed away from the source region} would transport momentum back to accelerate the zonal winds, which leads to a conversion from K$_{\rm E}$ to K$_{\rm M}$; the shear instability of the zonal-mean winds would generate eddies, which leads to a conversion from K$_{\rm M}$ to K$_{\rm E}$ (Figure~\ref{fig:LEC_framework}(e)). 
    \item[$\bullet$]Conversion rate from P$_{\rm M}$ to K$_{\rm M}$
    \begin{eqnarray}\label{eq:C(PM,KM)}
        {\rm C\left(P_M,K_M\right)}=-\int g\left[\bar{v}\right]\frac{\partial\left[\bar{Z}\right]}{a\partial\phi}dm,
    \end{eqnarray}
    which depends on the same processes as $\rm C(P_E,K_E)$ but in the large-scale circulation.
\end{enumerate}   

We do not calculate the generation rates of potential energy ($\rm G(P_{M})$ and $\rm G(P_{E})$) and the dissipation rates of kinetic energy ($\rm D(K_{M})$ and $\rm D(K_{E})$). This is because it is hard to list all diabatic heating, cooling, and damping processes from the reanalysis data and the GCMs' outputs. 

The formulas of $\rm C(P_M, K_M)$ and $\rm C(P_E, K_E)$ are so-called `v$\cdot$grad z' form. There is another form called `$\omega\cdot\alpha$' ($\alpha$ is specific volume), i.e., $\left[\bar{\omega}\right]\left[\bar{\alpha}\right]$ and $\left[\overline{\omega^{'}\alpha^{'}}+\bar{\omega}^{*}\bar{\alpha}^{*}\right]$, respectively. The `$\omega\cdot\alpha$' form is associated with different physical aspect of the conversion, that is, vertical motions of warm and cold air. The two forms may display different spatial distributions, but lead to the same global-mean vertical integrals \citep{Peixoto_1992,Marques_2009,Kim_2013}. Thus, different forms have tiny impacts on our conclusions. We use the `v$\cdot$grad z' form in this study, because wind velocities and geopotential height can be directly read from the reanalysis data and GCMs' outputs.

\end{appendix}

\begin{acknowledgments}
We are grateful to Eric Wolf for the release of the model ExoCAM and to D.B. Koll for the release of the changing-coordinates codes and helpful discussions. J.Y. acknowledges support from the National Natural Science Foundation of China (NSFC) under grants 42161144011 and 42075046.
  
The model ExoCAM is released at \url{https://github.com/storyofthewolf/ExoCAM}. The transformation codes for tidally locked coordinates are released at \url{https://github.com/ddbkoll/tidally locked-coordinates}. The calculation codes for the LEC are released at \url{https://doi.org/10.5281/zenodo.7472396}. The simulation data in this study are archived at \url{https://doi.org/10.5281/zenodo.7476074}. 
\end{acknowledgments}

\bibliography{sample631}{}

\begin{thebibliography}{}
\expandafter\ifx\csname natexlab\endcsname\relax\def\natexlab#1{#1}\fi
\providecommand{\url}[1]{\href{#1}{#1}}
\providecommand{\dodoi}[1]{doi:~\href{http://doi.org/#1}{\nolinkurl{#1}}}
\providecommand{\doeprint}[1]{\href{http://ascl.net/#1}{\nolinkurl{http://ascl.net/#1}}}
\providecommand{\doarXiv}[1]{\href{https://arxiv.org/abs/#1}{\nolinkurl{https://arxiv.org/abs/#1}}}

\bibitem[{Carone {et~al.}(2015)Carone, Keppens, \& Decin}]{Carone_2015}
Carone, L., Keppens, R., \& Decin, L. 2015, Monthly Notices of the Royal
  Astronomical Society, 453, 2412, \dodoi{10.1093/mnras/stv1752}

\bibitem[{Carone {et~al.}(2016)Carone, Keppens, \& Decin}]{Carone_2016}
---. 2016, Monthly Notices of the Royal Astronomical Society, 461, 1981,
  \dodoi{10.1093/mnras/stw1265}

\bibitem[{Debras {et~al.}(2020)Debras, Mayne, Baraffe, Jaupart, Mourier, Laibe,
  Goffrey, \& Thuburn}]{Debras_2020}
Debras, F., Mayne, N., Baraffe, I., {et~al.} 2020, A\&A, 633, A2,
  \dodoi{10.1051/0004-6361/201936110}

\bibitem[{{Del Genio} {et~al.}(1993){Del Genio}, Zhou, \&
  Eichler}]{DelGenio_1993}
{Del Genio}, A.~D., Zhou, W., \& Eichler, T.~P. 1993, Icarus, 101, 1,
  \dodoi{https://doi.org/10.1006/icar.1993.1001}

\bibitem[{Ding \& Pierrehumbert(2020)}]{Ding_2020b}
Ding, F., \& Pierrehumbert, R.~T. 2020, \apj, 901, L33,
  \dodoi{10.3847/2041-8213/abb941}

\bibitem[{Ding \& Wordsworth(2019)}]{Ding_2019}
Ding, F., \& Wordsworth, R.~D. 2019, \apj, 878, 117,
  \dodoi{10.3847/1538-4357/ab204f}

\bibitem[{Ding \& Wordsworth(2020)}]{Ding_2020a}
---. 2020, \apj, 891, L18, \dodoi{10.3847/2041-8213/ab77d1}

\bibitem[{Ding \& Wordsworth(2021)}]{Ding_2021}
---. 2021, The Planetary Science Journal, 2, 201, \dodoi{10.3847/psj/ac2236}

\bibitem[{Duan \& Wu(2005)}]{Duan_2005}
Duan, A., \& Wu, G. 2005, Science in China Series D: Earth Sciences, 48, 1293,
  \dodoi{10.1360/04yd0042}

\bibitem[{Edson {et~al.}(2011)Edson, Lee, Bannon, Kasting, \&
  Pollard}]{Edson_2011}
Edson, A., Lee, S., Bannon, P., Kasting, J.~F., \& Pollard, D. 2011, Icarus,
  212, 1, \dodoi{https://doi.org/10.1016/j.icarus.2010.11.023}

\bibitem[{Gill(1980)}]{Gill_1980}
Gill, A.~E. 1980, Q. J. R. Meteorol. Soc., 106, 447,
  \dodoi{10.1002/qj.49710644905}

\bibitem[{Hammond \& Lewis(2021)}]{Hammond_2021}
Hammond, M., \& Lewis, N.~T. 2021, Proceedings of the National Academy of
  Sciences, 118, \dodoi{10.1073/pnas.2022705118}

\bibitem[{Hammond \& Pierrehumbert(2018)}]{Hammond_2018}
Hammond, M., \& Pierrehumbert, R.~T. 2018, \apj, 869, 65,
  \dodoi{10.3847/1538-4357/aaec03}

\bibitem[{Hammond {et~al.}(2020)Hammond, Tsai, \& Pierrehumbert}]{Hammond_2020}
Hammond, M., Tsai, S.-M., \& Pierrehumbert, R.~T. 2020, \apj, 901, 78,
  \dodoi{10.3847/1538-4357/abb08b}

\bibitem[{Haqq-Misra {et~al.}(2018)Haqq-Misra, Wolf, Joshi, Zhang, \&
  Kopparapu}]{Haqq_Misra_2018}
Haqq-Misra, J., Wolf, E.~T., Joshi, M., Zhang, X., \& Kopparapu, R.~K. 2018,
  \apj, 852, 67, \dodoi{10.3847/1538-4357/aa9f1f}

\bibitem[{Harnik \& Chang(2003)}]{Harnik_2003}
Harnik, N., \& Chang, E. K.~M. 2003, Journal of Climate, 16, 480 ,
  \dodoi{10.1175/1520-0442(2003)016<0480:STVASI>2.0.CO;2}

\bibitem[{Hernández-Deckers \& von Storch(2010)}]{Deckers_2010}
Hernández-Deckers, D., \& von Storch, J.-S. 2010, Journal of Climate, 23,
  3874, \dodoi{10.1175/2010JCLI3176.1}

\bibitem[{Holton \& Hakim(2013)}]{Holton_2013}
Holton, J.~R., \& Hakim, G.~J. 2013, An introduction to Dynamic Meteorology:
  5th (London: Academic Press.),
  \dodoi{https://doi.org/10.1016/C2009-0-63394-8}

\bibitem[{{Ingersoll}(2013)}]{Ingersoll_2013}
{Ingersoll}, A.~P. 2013, Planetary Climates (New Jersey: Princeton University
  Press)

\bibitem[{Joshi {et~al.}(1997)Joshi, Haberle, \& Reynolds}]{Joshi_1997}
Joshi, M., Haberle, R., \& Reynolds, R. 1997, Icarus, 129, 450 ,
  \dodoi{https://doi.org/10.1006/icar.1997.5793}

\bibitem[{Joshi {et~al.}(2020)Joshi, Elvidge, Wordsworth, \&
  Sergeev}]{Joshi_2020}
Joshi, M.~M., Elvidge, A.~D., Wordsworth, R., \& Sergeev, D. 2020, The
  Astrophysical Journal Letters, 892, L33, \dodoi{10.3847/2041-8213/ab7fb3}

\bibitem[{Kanno \& Iwasaki(2022)}]{Yuki_2022}
Kanno, Y., \& Iwasaki, T. 2022, Journal of Geophysical Research: Atmospheres,
  127, e2021JD036380, \dodoi{https://doi.org/10.1029/2021JD036380}

\bibitem[{Kim \& Choi(2017)}]{Kim_2017}
Kim, W., \& Choi, Y.-S. 2017, Climate Dynamics, 49, 3605,
  \dodoi{10.1007/s00382-017-3533-0}

\bibitem[{Kim \& Kim(2013)}]{Kim_2013}
Kim, Y.-H., \& Kim, M.-K. 2013, Climate Dynamics, 40, 1499,
  \dodoi{10.1007/s00382-012-1358-4}

\bibitem[{Kite {et~al.}(2011)Kite, Gaidos, \& Manga}]{Kite_2011}
Kite, E.~S., Gaidos, E., \& Manga, M. 2011, \apj, 743, 41,
  \dodoi{10.1088/0004-637x/743/1/41}

\bibitem[{Koll \& Abbot(2015)}]{Koll_2015}
Koll, D. D.~B., \& Abbot, D.~S. 2015, \apj, 802, 21,
  \dodoi{10.1088/0004-637x/802/1/21}

\bibitem[{Koll \& Abbot(2016)}]{Koll_2016}
---. 2016, \apj, 825, 99, \dodoi{10.3847/0004-637x/825/2/99}

\bibitem[{Komacek {et~al.}(2019)Komacek, Jansen, Wolf, \& Abbot}]{Komacek_2019}
Komacek, T.~D., Jansen, M.~F., Wolf, E.~T., \& Abbot, D.~S. 2019, \apj, 883,
  46, \dodoi{10.3847/1538-4357/ab3980}

\bibitem[{Kundu {et~al.}(2016)Kundu, Cohen, \& Dowling}]{Kundu_fluid}
Kundu, P.~K., Cohen, I.~M., \& Dowling, D.~R. 2016, in Fluid Mechanics (Sixth
  Edition), sixth edition edn., ed. P.~K. Kundu, I.~M. Cohen, \& D.~R. Dowling
  (Boston: Academic Press), 49--76,
  \dodoi{https://doi.org/10.1016/B978-0-12-405935-1.00002-2}

\bibitem[{Leconte {et~al.}(2013)Leconte, Forget, Charnay, Wordsworth, Selsis,
  Millour, \& Spiga}]{Leconte_2013}
Leconte, J., Forget, F., Charnay, B., {et~al.} 2013, A\&A, 554, A69,
  \dodoi{10.1051/0004-6361/201321042}

\bibitem[{Lee \& Richardson(2010)}]{Lee_2010}
Lee, C., \& Richardson, M.~I. 2010, Journal of Geophysical Research: Planets,
  115, \dodoi{https://doi.org/10.1029/2009JE003490}

\bibitem[{Li {et~al.}(2007)Li, Ingersoll, Jiang, Feldman, \& Yung}]{Li_2007}
Li, L., Ingersoll, A.~P., Jiang, X., Feldman, D., \& Yung, Y.~L. 2007,
  Geophysical Research Letters, 34,
  \dodoi{https://doi.org/10.1029/2007GL029985}

\bibitem[{Lorenz(1955)}]{Lorenz_1955}
Lorenz, E.~N. 1955, Tellus, 7, 157, \dodoi{10.3402/tellusa.v7i2.8796}

\bibitem[{Marques {et~al.}(2010)Marques, Rocha, \& Corte-Real}]{Marques_2010}
Marques, C., Rocha, A., \& Corte-Real, J. 2010, Dynamics of Atmospheres and
  Oceans, 50, 375, \dodoi{https://doi.org/10.1016/j.dynatmoce.2010.03.003}

\bibitem[{Marques {et~al.}(2011)Marques, Rocha, \& Corte-Real}]{Marques_2011}
Marques, C. A.~F., Rocha, A., \& Corte-Real, J. 2011, Climate Dynamics, 36,
  1767, \dodoi{10.1007/s00382-010-0828-9}

\bibitem[{Marques {et~al.}(2009)Marques, Rocha, Corte-Real, Castanheira,
  Ferreira, \& Melo-Gonçalves}]{Marques_2009}
Marques, C. A.~F., Rocha, A., Corte-Real, J., {et~al.} 2009, International
  Journal of Climatology, 29, 159, \dodoi{https://doi.org/10.1002/joc.1704}

\bibitem[{Matsuno(1966)}]{Matsuno_1966}
Matsuno, T. 1966, J. Meteorol. Soc. Japan, 44, 25,
  \dodoi{10.2151/jmsj1965.44.1_25}

\bibitem[{Mendonça(2019)}]{Mendonca_2020}
Mendonça, J.~M. 2019, Monthly Notices of the Royal Astronomical Society, 491,
  1456, \dodoi{10.1093/mnras/stz3050}

\bibitem[{Merlis \& Schneider(2010)}]{Merlis_2010}
Merlis, T.~M., \& Schneider, T. 2010, Journal of Advances in Modeling Earth
  Systems, 2, \dodoi{https://doi.org/10.3894/JAMES.2010.2.13}

\bibitem[{Michaelides(2021)}]{Michaelides_2021}
Michaelides, S. 2021, Climate, 9, \dodoi{10.3390/cli9120180}

\bibitem[{Noda {et~al.}(2017)Noda, Ishiwatari, Nakajima, Takahashi, Takehiro,
  Onishi, Hashimoto, Kuramoto, \& Hayashi}]{Noda_2017}
Noda, S., Ishiwatari, M., Nakajima, K., {et~al.} 2017, Icarus, 282, 1,
  \dodoi{https://doi.org/10.1016/j.icarus.2016.09.004}

\bibitem[{Olbers {et~al.}(2012)Olbers, Willebrand, \& Eden}]{Olbers_2012}
Olbers, D., Willebrand, J., \& Eden, C. 2012, Ocean Dynamics (Berlin,
  Heidelberg: Springer), \dodoi{https://doi.org/10.1007/978-3-642-23450-7}

\bibitem[{Pan {et~al.}(2017)Pan, Li, Jiang, Li, Zhang, Wang, \&
  Ingersoll}]{Pan_2017}
Pan, Y., Li, L., Jiang, X., {et~al.} 2017, Nature Communications, 8, 14367,
  \dodoi{10.1038/ncomms14367}

\bibitem[{Peixóto \& Oort(1974)}]{Peixoto_1974}
Peixóto, J.~P., \& Oort, A.~H. 1974, Journal of Geophysical Research
  (1896-1977), 79, 2149, \dodoi{https://doi.org/10.1029/JC079i015p02149}

\bibitem[{Peixóto \& Oort(1992)}]{Peixoto_1992}
---. 1992, Physics of Climate (New York: American Institute of Physics)

\bibitem[{Perez-Becker \& Showman(2013)}]{Perez_Becker_2013}
Perez-Becker, D., \& Showman, A.~P. 2013, \apj, 776, 134,
  \dodoi{10.1088/0004-637x/776/2/134}

\bibitem[{Pierrehumbert(2010)}]{Pierrehumbert_2010}
Pierrehumbert, R.~T. 2010, \apj, 726, L8, \dodoi{10.1088/2041-8205/726/1/l8}

\bibitem[{Pierrehumbert \& Hammond(2019)}]{Pierrehumbert_Hammond_2019}
Pierrehumbert, R.~T., \& Hammond, M. 2019, Annual Review of Fluid Mechanics,
  51, 275, \dodoi{10.1146/annurev-fluid-010518-040516}

\bibitem[{Sergeev {et~al.}(2022)Sergeev, Fauchez, Turbet, Boutle, Tsigaridis,
  Way, Wolf, Domagal-Goldman, Forget, Haqq-Misra, Kopparapu, Lambert, Manners,
  \& Mayne}]{Sergeev_2022}
Sergeev, D.~E., Fauchez, T.~J., Turbet, M., {et~al.} 2022, The Planetary
  Science Journal, 3, 212, \dodoi{10.3847/PSJ/ac6cf2}

\bibitem[{Showman {et~al.}(2009)Showman, Cho, \& Menou}]{Showman_2009}
Showman, A.~P., Cho, J. Y.-K., \& Menou, K. 2009, Atmospheric Circulation of
  Exoplanets, \dodoi{10.48550/arXiv.0911.3170}

\bibitem[{Showman \& Polvani(2010)}]{Showman_2010}
Showman, A.~P., \& Polvani, L.~M. 2010, Geophys. Res. Lett, 37,
  \dodoi{10.1029/2010GL044343}

\bibitem[{Showman \& Polvani(2011)}]{Showman_2011}
---. 2011, \apj, 738, 71, \dodoi{10.1088/0004-637x/738/1/71}

\bibitem[{Song \& Yang(2021)}]{Song_2021}
Song, X., \& Yang, J. 2021, Frontiers in Astronomy and Space Sciences, 8,
  \dodoi{10.3389/fspas.2021.708023}

\bibitem[{Trenberth(1991)}]{Trenberth_1991}
Trenberth, K.~E. 1991, Journal of Atmospheric Sciences, 48, 2159 ,
  \dodoi{10.1175/1520-0469(1991)048<2159:STITSH>2.0.CO;2}

\bibitem[{Tsai {et~al.}(2014)Tsai, Dobbs-Dixon, \& Gu}]{Tsai_2014}
Tsai, S.-M., Dobbs-Dixon, I., \& Gu, P.-G. 2014, \apj, 793, 141,
  \dodoi{10.1088/0004-637x/793/2/141}

\bibitem[{Turbet {et~al.}(2022)Turbet, Fauchez, Sergeev, Boutle, Tsigaridis,
  Way, Wolf, Domagal-Goldman, Forget, Haqq-Misra, Kopparapu, Lambert, Manners,
  Mayne, \& Sohl}]{Turbet_2022}
Turbet, M., Fauchez, T.~J., Sergeev, D.~E., {et~al.} 2022, The Planetary
  Science Journal, 3, 211, \dodoi{10.3847/PSJ/ac6cf0}

\bibitem[{Ulbrich \& Speth(1991)}]{Ulbrich_1991}
Ulbrich, U., \& Speth, P. 1991, Meteorology and Atmospheric Physics, 45, 125,
  \dodoi{10.1007/BF01029650}

\bibitem[{Vallis(2019)}]{Vallis_2019}
Vallis, G.~K. 2019, Essentials of atmospheric and oceanic dynamics (Cambridge,
  U.K.: Cambridge University Press), \dodoi{10.1017/9781107588431}

\bibitem[{von Storch {et~al.}(2012)von Storch, Eden, Fast, Haak,
  Hernández-Deckers, Maier-Reimer, Marotzke, \& Stammer}]{Storch_2012}
von Storch, J.-S., Eden, C., Fast, I., {et~al.} 2012, Journal of Physical
  Oceanography, 42, 2185, \dodoi{10.1175/JPO-D-12-079.1}

\bibitem[{Wang \& Yang(2021)}]{Wang_2020}
Wang, S., \& Yang, J. 2021, \apj, 907, 28, \dodoi{10.3847/1538-4357/abcf2a}

\bibitem[{Wang \& Yang(2022)}]{Wang_2022}
---. 2022, The Planetary Science Journal, 3, 171, \dodoi{10.3847/psj/ac6d65}

\bibitem[{Wolf \& Toon(2014)}]{Wolf_2014}
Wolf, E., \& Toon, O. 2014, Astrobiology, 14, 241,
  \dodoi{10.1089/ast.2013.1112}

\bibitem[{Wolf {et~al.}(2022)Wolf, Kopparapu, Haqq-Misra, \&
  Fauchez}]{Wolf_2022}
Wolf, E.~T., Kopparapu, R., Haqq-Misra, J., \& Fauchez, T.~J. 2022, The
  Planetary Science Journal, 3, 7, \dodoi{10.3847/psj/ac3f3d}

\bibitem[{Wolf {et~al.}(2017)Wolf, Shields, Kopparapu, Haqq-Misra, \&
  Toon}]{Wolf_et_al_2017}
Wolf, E.~T., Shields, A.~L., Kopparapu, R.~K., Haqq-Misra, J., \& Toon, O.~B.
  2017, \apj, 837, 107, \dodoi{10.3847/1538-4357/aa5ffc}

\bibitem[{Wolf \& Toon(2015)}]{Wolf_2015}
Wolf, E.~T., \& Toon, O.~B. 2015, Journal of Geophysical Research: Atmospheres,
  120, 5775, \dodoi{https://doi.org/10.1002/2015JD023302}

\bibitem[{Wordsworth(2015)}]{Wordsworth_2015}
Wordsworth, R. 2015, \apj, 806, 180, \dodoi{10.1088/0004-637x/806/2/180}

\bibitem[{Yamamoto \& Takahashi(2006)}]{Yamamoto_2006}
Yamamoto, M., \& Takahashi, M. 2006, Journal of the Atmospheric Sciences, 63,
  3296, \dodoi{10.1175/JAS3859.1}

\bibitem[{Yang {et~al.}(2014)Yang, Boué, Fabrycky, \& Abbot}]{Yang_2014}
Yang, J., Boué, G., Fabrycky, D.~C., \& Abbot, D.~S. 2014, \apj, 787, L2,
  \dodoi{10.1088/2041-8205/787/1/L2}

\bibitem[{Yang {et~al.}(2013)Yang, Cowan, \& Abbot}]{Yang_2013}
Yang, J., Cowan, N.~B., \& Abbot, D.~S. 2013, \apj, 771, L45,
  \dodoi{10.1088/2041-8205/771/2/l45}

\bibitem[{Zhang \& Yang(2020)}]{Zhang_2020}
Zhang, Y., \& Yang, J. 2020, \apj, 901, L36, \dodoi{10.3847/2041-8213/abb87f}

\end{thebibliography}
\bibliographystyle{aasjournal}



\end{document}